\documentclass[sigplan,10pt,authorversion]{acmart}

\AtBeginDocument{%
  \providecommand\BibTeX{{%
    \normalfont B\kern-0.5em{\scshape i\kern-0.25em b}\kern-0.8em\TeX}}}

\setcopyright{rightsretained}
\copyrightyear{2024}
\acmDOI{}

\acmConference[Under review]{}{do not distribute}{the authors.}
\acmBooktitle{} 
\usepackage{placeins}
\usepackage{graphicx}
\usepackage{booktabs} % for professional tables
\usepackage{siunitx}
\usepackage{amsmath}
\usepackage[tableposition=top]{caption}
\usepackage{subcaption}
\usepackage[most]{tcolorbox}
\usepackage{soul}
\usepackage{array}
\newtcolorbox{push}{colback=blue!10!white,boxrule=0pt, top=0pt,bottom=0pt, left=0.1pt}
\newtcolorbox{pull}{colback=red!10!white,boxrule=0pt,top=0pt,bottom=0pt, left=0.1pt}
\newtcolorbox{gen}{colback=green!10!white,boxrule=0pt, top=0pt,bottom=0pt, left=0.1pt}
\usepackage{hyperref}
\usepackage[capitalise]{cleveref}
\usepackage[noend]{algorithmic}
\usepackage{setspace}
\usepackage{xspace}
\usepackage{enumitem}
\usepackage{multirow}

\newcommand{\pollen}{\textit{Pollen}\xspace}
\newcommand{\flower}{\textit{Flower}\xspace}
\newcommand{\fedscale}{\textit{FedScale}\xspace}
\newcommand{\flute}{\textit{Flute}\xspace}
\newcommand{\parrot}{\textit{Parrot}\xspace}
\newcommand{\pfl}{\textit{pfl}\xspace}

\newcolumntype{T}[1]{%
    >{\centering\arraybackslash\hspace{0pt}}p{#1}}%
\newcolumntype{F}[1]{%
    >{\centering\arraybackslash\hspace{0pt}}c{#1}}%

\begin{document}

\title{\pollen: High-throughput Federated Learning Simulation via Resource-Aware Client Placement}

\author{Lorenzo Sani}
\authornote{Authors contributed equally to this research.}
\email{ls985@cam.ac.uk}
\orcid{}
\affiliation{%
  \institution{University of Cambridge}
  % \streetaddress{P.O. Box 1212}
  \city{Cambridge}
  \country{United Kingdom}
}
\author{Pedro Porto Buarque de Gusmão}
\authornotemark[1]
\email{p.gusmao@surrey.ac.uk}
\affiliation{%
  \institution{University of Surrey}
  % \streetaddress{P.O. Box 1212}
  \city{Guildford}
  \country{United Kingdom}
}

\author{Alex Iacob}
\authornotemark[1]
\email{aai30@cam.ac.uk}
\affiliation{%
  \institution{University of Cambridge}
  % \streetaddress{1 Th{\o}rv{\"a}ld Circle}
  \city{Cambridge}
  \country{United Kingdom}
}

\author{Wanru Zhao}
\email{wz341@cam.ac.uk}
\affiliation{%
  \institution{University of Cambridge}
  \city{Cambridge}
  \country{United Kingdom}
}

\author{Xinchi Qiu}
\email{xq227@cam.ac.uk}
\affiliation{%
  \institution{University of Cambridge}
  \city{Cambridge}
  \country{United Kingdom}
}

\author{Yan Gao}
\email{yg381@cam.ac.uk}
\affiliation{%
  \institution{University of Cambridge}
  % \streetaddress{30 Shuangqing Rd}
  \city{Cambridge}
  % \state{Beijing Shi}
  \country{United Kingdom}
}

\author{Javier Fernandez-Marques}
\email{javier@flower.ai}
\affiliation{%
  \institution{Flower Labs}
  % \streetaddress{8600 Datapoint Drive}
  \city{Cambridge}
  % \state{Texas}
  \country{United Kingdom}
  % \postcode{78229}
}

\author{Nicholas Donald Lane}
\email{ndl32@cam.ac.uk}
\affiliation{%
  \institution{University of Cambridge, Flower Labs}
  % \streetaddress{1 Th{\o}rv{\"a}ld Circle}
  \city{Cambridge}
  \country{United Kingdom}
}

\renewcommand{\shortauthors}{Sani, et al.}

\begin{abstract}
  Federated Learning~(FL) is a privacy-focused machine learning paradigm that collaboratively trains models directly on edge devices. Simulation plays an essential role in FL adoption, helping develop novel aggregation and client sampling strategies. However, current simulators cannot emulate large-scale systems in a time-efficient manner, which limits their utility and casts doubts on generalizability.

This work proposes \pollen, a novel resource-aware system for speeding up simulations.
\pollen addresses two limiting factors from existing simulators: (a) communication inefficiency derived from pull-based client execution and (b) inadequate load balance when using heterogeneous hardware.
\pollen executes high-throughput FL simulations at scale by (a) using a push-based client placement system, (b) learning how an adaptable scheduling of clients based on hardware statistics (c) estimating the optimal number of concurrent workers per GPU.
We evaluate \pollen on four representative FL tasks and show that \pollen's placement model increases GPU utilization and reduces idle time.
We compare \pollen to \flower, \flute, \fedscale,
\parrot, and \pfl and show experimental speed-ups of days or weeks.
\end{abstract}

\begin{CCSXML}
<ccs2012>
   <concept>
       <concept_id>10010147.10010178</concept_id>
       <concept_desc>Computing methodologies~Artificial intelligence</concept_desc>
       <concept_significance>500</concept_significance>
       </concept>
   <concept>
       <concept_id>10010147.10010178.10010199</concept_id>
       <concept_desc>Computing methodologies~Planning and scheduling</concept_desc>
       <concept_significance>500</concept_significance>
       </concept>
   <concept>
       <concept_id>10010147.10010341.10010366.10010369</concept_id>
       <concept_desc>Computing methodologies~Simulation tools</concept_desc>
       <concept_significance>500</concept_significance>
       </concept>
 </ccs2012>
\end{CCSXML}

\ccsdesc[500]{Computing methodologies~Artificial intelligence}
\ccsdesc[500]{Computing methodologies~Planning and scheduling}
\ccsdesc[500]{Computing methodologies~Simulation tools}

% \keywords{Do, Not, Us, This, Code, Put, the, Correct, Terms, for,
\keywords{federated learning, simulation systems, machine learning}



\settopmatter{printfolios=false,printccs=false,printacmref=false}\maketitle
\pagestyle{plain}
% \vspace{-0.4cm}
\section{Introduction}\label{sec:intro}
Federated learning (FL)~\citep{OgFedAvg} is a privacy-preserving machine learning paradigm where clients train a prediction model without sharing local data.
Despite its potential to harvest data and computing power from millions of devices, we know surprisingly little about how well FL solutions may scale to such numbers. Having the right tools to run simulations over large cohorts efficiently would allow us to investigate the challenges only noticeable at those scales ~\citep{AdvancedAndOpenProblems}. %\wz{as follows:}. 

\begin{figure}[t]
    \centering
    \includegraphics[height=0.7\columnwidth]{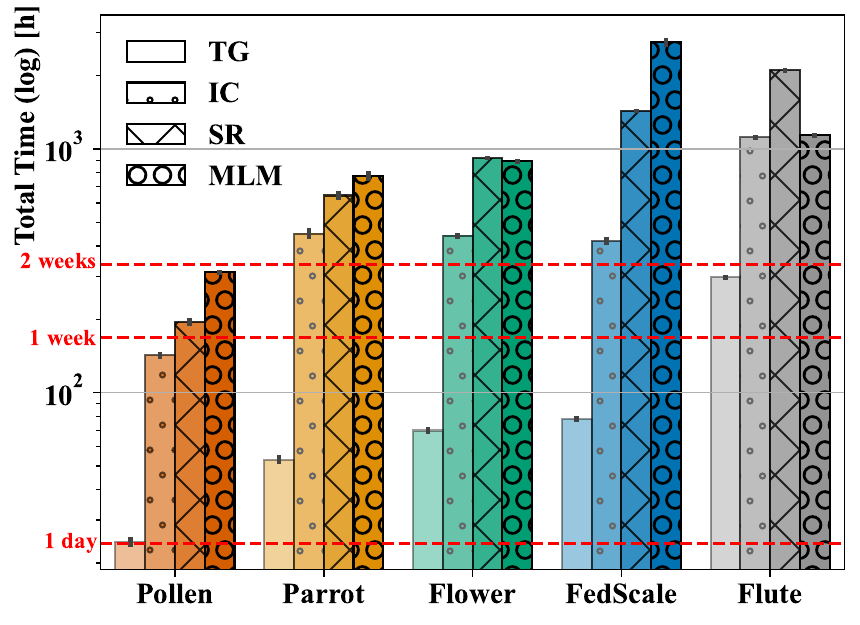}
    % \captionsetup{font=small,labelfont=bf}
    \caption{For large-scale experiments of $5$ million clients trained across \num{5000} rounds, \pollen makes possible previously unfeasible experiments and outperforms all other frameworks. For example, on the \emph{Image Classification} task \pollen executes in less than one week, outmatching the two-week training time of its competitors. \cref{sec:experimental_design} describes the design of this experiment. We also provide a comparison against the recently released~\footref{foot:pfl_release} \pfl~\citep{pfl_paper} framework in \cref{sec:pfl_comparison}. }
    \label{fig:total_training_time_large_scale}
    \Description[Figure 1]{For large-scale experiments of $5$ million clients trained across \num{5000} rounds, \pollen makes possible previously unfeasible experiments and outperforms all other frameworks. For example, on the \emph{Image Classification} task \pollen executes in less than one week, outmatching the two-week training time of its competitors. \cref{sec:experimental_design} describes the design of this experiment. We also provide a comparison against the recently released~\footref{foot:pfl_release} \pfl~\citep{pfl_paper} framework in \cref{sec:pfl_comparison}.}
% \vspace{-0.6cm}
\end{figure}

%\wz{Or adding a topic sentence to explain that the following several aspects listed is limited by the lack of a large-scale simulation platform.} 
Out of the several open problems in the area of large-scale federated learning, the following require extensive and resource-demanding experimentation.
As cohort size increases, differences in hardware may dictate which clients may time out, raising an issue of fairness.
Large cohorts also make the effects of client data heterogeneity more pronounced, making training a single global model difficult even when using state-of-the-art (SOTA) techniques.
Finally, when considering world-scale experiments, client availability and connectivity patterns will directly impact a model's performance, proportional to the training population and geographical distribution. 

Although existing FL frameworks, such as \flower \citep{Flower}, \fedscale \citep{FedScale}, \flute \citep{Flute} and \pfl \cite{pfl_paper}, provide support for simulation, their architectural designs have three primary limitations.
First, they use a communication-inefficient design, where GPU workers pull clients from a server queue and send back a model for each client.
Second, these frameworks do not account for hardware diversity, leading to a straggler effect when training on multi-GPUs. 
Third, most implementations fail to exploit the high bandwidth available between the GPUs of one machine and instead rely on communication-inefficient methods.
Fourth, they have no means of estimating the optimal number of worker processes that should be involved in the execution, either wasting resources or requiring heavy manual tuning.

These limitations may go unnoticed at small/medium scales as they have a restricted impact on wall clock time; however, they make large-scale simulations impractical or unfeasible as experiments may take weeks or even months to complete. 
In this work, we propose \pollen as a general-purpose FL simulator capable of achieving significant speed-ups over existing systems by addressing the abovementioned issues.
\pollen overcomes previous frameworks' shortcomings with a novel design that minimizes inter-machine communication and maximizes resource efficiency.
Additionally, it can exploit GPUs' whole computing power by leveraging a novel concurrency-aware client placement model.

While this design is highly communication efficient, we show that naive client allocation can nevertheless still result in straggling GPUs, decreasing GPU utilization, and increasing the system's idle time. Thus, for every round \pollen uses a concurrency-aware machine learning model to one-shot allocate clients onto each GPU based on the number of batches held by a client, the speed of the GPU, and the estimated optimal number of worker processes for that GPU given its VRAM and the characteristics of the system it is embedded in. In summary, our main contributions are:

\begin{enumerate}
    \item We build \pollen as a high-performance FL experimentation system capable of executing tens of thousands of clients per round from populations of millions without manual tuning for the specific hardware available. \Cref{fig:total_training_time_large_scale} shows that \pollen can achieve significant reductions in execution time over the simulation systems used by \flower, \fedscale, \flute, and \parrot for large-scale experiments.
    Furthermore, \cref{sec:pfl_comparison} shows that \pollen outperforms the recently released\footnote{ \citet{pfl_paper} was released \num{10} days before the SOSP deadline.\label{foot:pfl_release}} \pfl~\citep{pfl_paper} on its two representative tasks. \pollen's design and placement model allow it to complete
    experiments that could take weeks, in mere days, enabling the investigation of realistic production-scale federated systems without prohibitive hardware acquisition costs. %\vspace{-0.5cm}
    \item We construct a detailed analysis of our placement model, built on three pillars, and independently verify the effectiveness of each component. These pillars are: (a) optimal concurrency estimation, (b) a robust log-linear model used to predict client training time, and (c) adaptive error correction for the model based on recent data. %\vspace{-0.5cm}
    \item Finally, we analyze the ability of \pollen to maintain maximal GPU utilization and minimize idle time during execution. Our comparison against standard placement strategies shows that misplacing clients on the GPUs results in systematically wasted compute time.
\end{enumerate}
\pollen is a drop-in replacement for the simulator of the \flower framework, supporting the same algorithms. Because \pollen offers a \textbf{sizable increase in FL simulation speed}, we have \textbf{beta-released it to AI labs} run by Samsung, Nokia, TCL, and Zenseact, and they have \textbf{run several thousand of workloads} with \pollen since it was released. We intend for it to receive a wide release as an open-source project to extend these benefits to the broader FL community.

\begin{figure}[ht]
    \centering
    \noindent\includegraphics[width=\columnwidth]{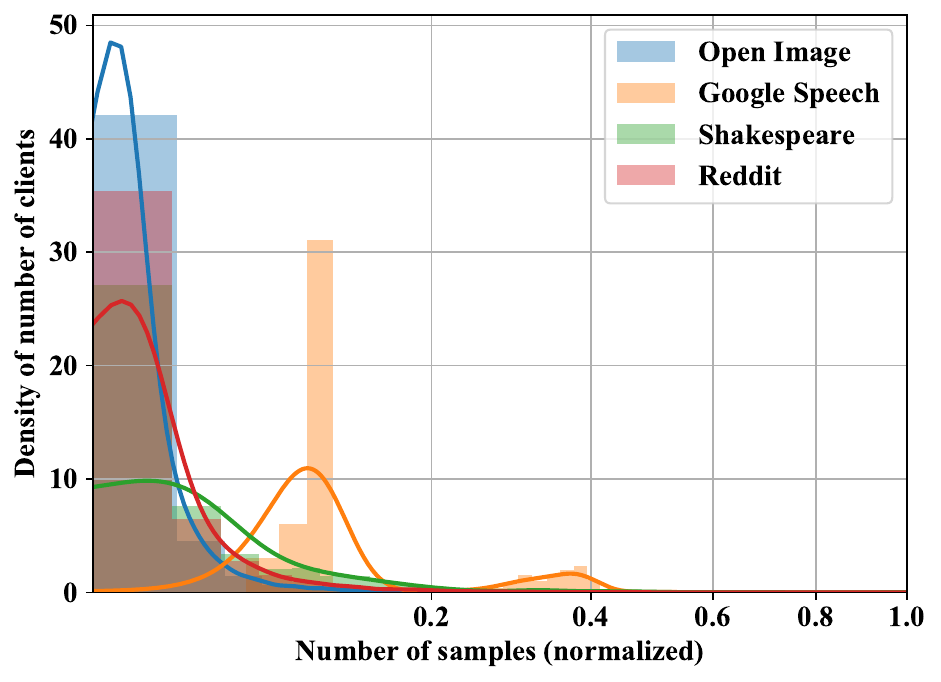}
    % \captionsetup{font=small,labelfont=bf}
    \caption{Dataset size distribution over clients for OpenImage, Google Speech, Shakespeare, and Reddit. The x-axis is non-linear, and the Reddit dataset is subsampled for comparison. Real-world federated datasets are naturally unbalanced, representing a challenge when optimizing resource utilization for machine learning algorithms.}
    \label{fig:client_heterogeneity}
    \Description[Figure 2]{Dataset size distribution over clients for OpenImage, Google Speech, Shakespeare, and Reddit. The x-axis is non-linear, and the Reddit dataset is subsampled for comparison. Real-world federated datasets are naturally unbalanced, representing a challenge when optimizing resource utilization for machine learning algorithms.}
% \vspace{-0.4cm}
\end{figure}

\section{Simulating Federated Learning}\label{sec:sim_fl}

In the following sections, we will argue that large-scale FL simulations open up relevant research directions and are absolutely necessary for the development of the field. 
First, we must consider the execution mode of Federated algorithms, such as Federated Averaging~\citep{OgFedAvg}, which is the most broadly accepted.
FL algorithms use a server to store the global model and synchronize devices across \emph{rounds}.
At the start of a round, the server sends the model to each client to train it on local data.
Then, clients send the models to the server for aggregation to create the next round's model.
In this work, we assume a fixed client participation rate and that the population size (number of clients in the federation) determines the scale of a federated experiment.

Recent research has shown there is a gap regarding large-scale federated settings.
\citet{LargeCohorts} show that FL algorithms cannot efficiently aggregate large client numbers due to their data-induced model dissimilarity.
The work of \citet{UnderstandingModelAveragingInFL} further indicates that the benefits of averaging more clients saturate, i.e.~alternative and more effective aggregation strategies are yet to be invented.
Furthermore, practical issues of coordinating clients with diverse hardware and activity patterns have not been fully overcome~\citep{AdvancedAndOpenProblems}.
These problems become crucial to federated systems in real-world production orchestrating the training of $10^{6}$ devices worldwide~\citep{ScaleAndSystemDesign}.
Crucially, it is currently unfeasible to research large-scale settings since FL frameworks are insufficiently optimized for simulation. Moreover, small/medium-scale experiments cannot provide sufficient information regarding large-scale scenarios. Our proposed system, \pollen, erases this limitation by reducing the execution time of \emph{large} simulations of common ML tasks from months to weeks when run on clusters containing multiple nodes with unbalanced GPUs~(see \cref{subsec:hard_conf}), as we show in \cref{sec:eval}.

\begin{figure}[t]
        \centering
    \noindent\includegraphics[height=0.7\columnwidth]{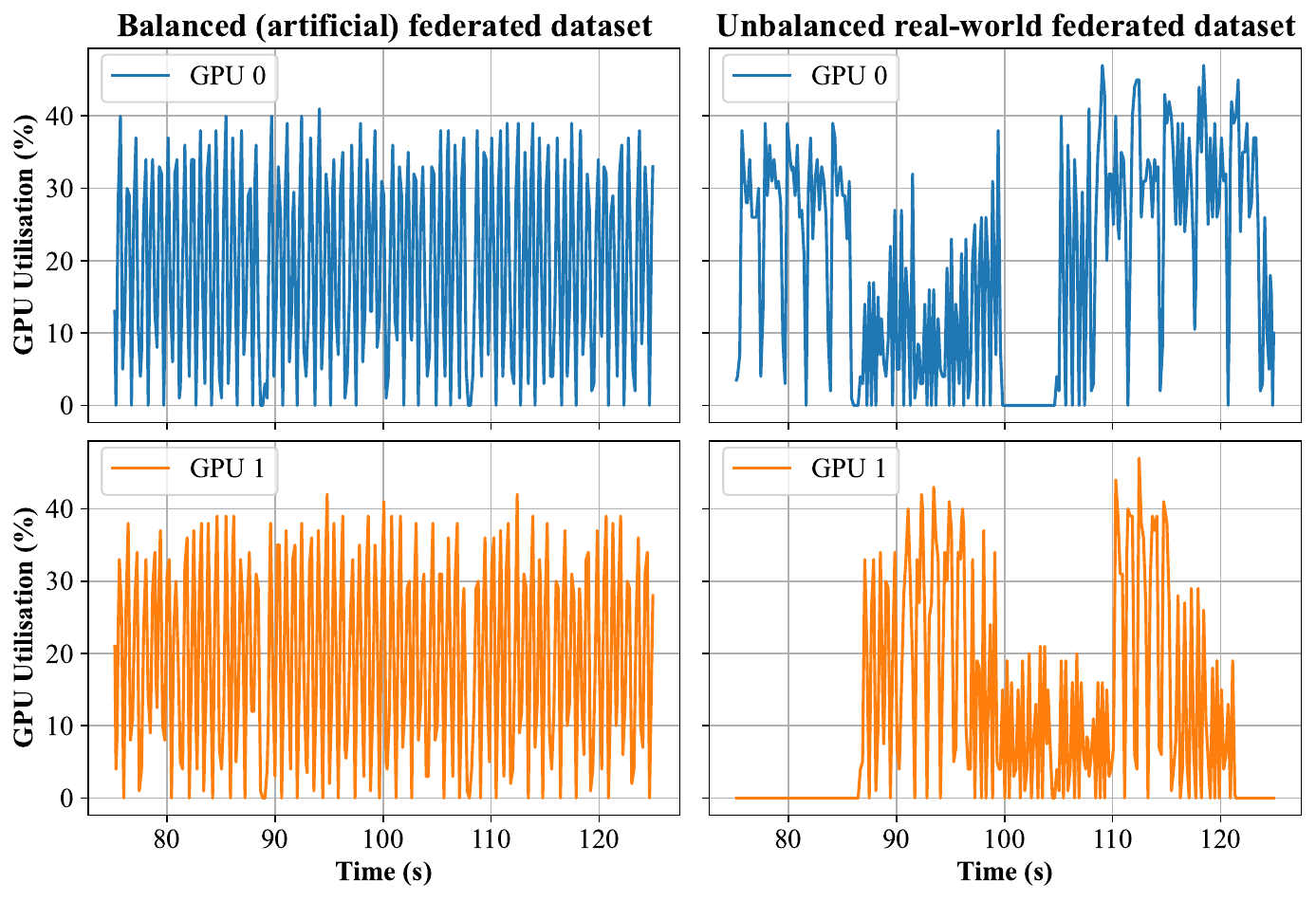}
        % \captionsetup{font=small,labelfont=bf}
    \caption{GPU utilization on two A40 GPUs training clients having the same number of samples~(left) versus clients having different numbers of samples following a naturally partitioned dataset, i.e.\ OpenImage ~(right). While the ideal scenario~(left) obtains balanced GPU idle times of $12.3$ and $12.6$ seconds, respectively, the naturally partitioned one results in $16.5$ and $40.5$ seconds, respectively. The comparison showcases the detrimental impact that unbalanced real-world datasets, such as those in \cref{fig:client_heterogeneity}, may have on GPU utilization.}
    \label{fig:GpuUtilComparison}
    \Description[Figure 3]{GPU utilization on two A40 GPUs training clients having the same number of samples~(left) versus clients having different numbers of samples following a naturally partitioned dataset, i.e.\ OpenImage ~(right). While the ideal scenario~(left) obtains balanced GPU idle times of $12.3$ and $12.6$ seconds, respectively, the naturally partitioned one results in $16.5$ and $40.5$ seconds, respectively. The comparison showcases the detrimental impact that unbalanced real-world datasets, such as those in \cref{fig:client_heterogeneity}, may have on GPU utilization.}
    % \vspace{-0.4cm}
\end{figure}

\subsection{Diverse Client Load}\label{subsec:client_heterogeneity}

In FL, clients produce data at vastly different rates~\cite{li2020federated}, resulting in unbalanced local datasets.
\Cref{fig:client_heterogeneity} shows client dataset size distributions for four standard naturally partitioned datasets.
While some frameworks circumvent this by fixing the number of local client training steps~(e.g., \fedscale), this assumes balanced datasets and leads to two issues that impact the outcome of the training procedure.

First, it limits the contribution of clients with large datasets.
Second, clients with small datasets must reuse data.
Because of these considerations, FL literature generally relies on a fixed number of local epochs, respecting client dataset size.
We refrain from fixing a constant number of steps and argue that this is not a reasonable assumption at a framework level.

Given their dataset's size, efficient simulations must account for clients' different training times.
We argue that the properties of federated datasets, together with the diversity of ML pre-processing pipelines, create the \emph{opportunity} to design \emph{adaptive placement strategies} which accommodate different scenarios.
Naively distributing clients across the available GPU devices used for simulation may result in idle GPU devices for repeated windows of time across the simulation. 
As shown by \cref{fig:GpuUtilComparison}, randomly distributing clients across GPUs for naturally partitioned datasets can significantly increase GPU idle times compared to the ideal scenario where all clients have the same number of samples. This has a large impact on GPU utilization; for example, the two A40s from \cref{fig:GpuUtilComparison} obtain a balanced average GPU utilization of $19\%$ in the ideal scenario while achieving $10.8\%$ and $19.7\%$ respectively in the naturally partitioned one. The data collected is a snapshot of a long-running \emph{Image Classification} task with \num{10} clients per round, covering \num{3} rounds in the middle of execution. 

\begin{figure}[t]
    \centering
    \noindent\includegraphics[width=0.95\columnwidth]{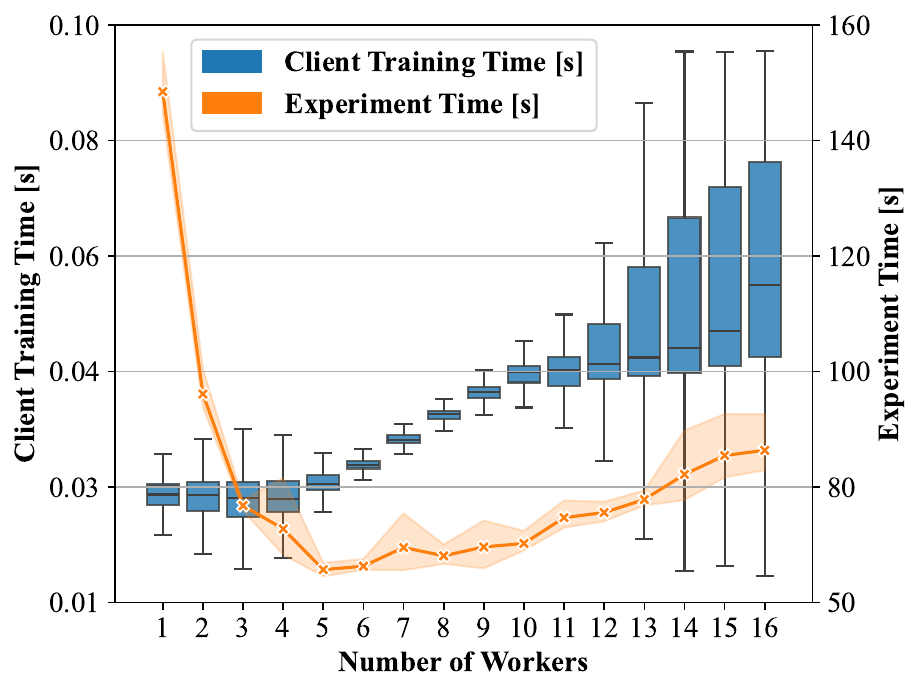}
    % \captionsetup{font=small,labelfont=bf}
    \caption{The number of worker processes has optimum depending on the hardware available and the task-specific workload. In this experiment, the same FL simulation is performed with a different number of worker processes. While the single-client training time increases proportionally to the number of workers, the total runtime of the experiment decreases until it reaches the optimum of 5 workers.}
    \label{fig:concurrency_optimum}
    \Description[Figure 4]{The number of worker processes has optimum depending on the hardware available and the task-specific workload. In this experiment, the same FL simulation is performed with a different number of worker processes. While the single-client training time increases proportionally to the number of workers, the total runtime of the experiment decreases until it reaches the optimum of 5 workers.}
% \vspace{-0.4cm}
\end{figure}

\subsection{Simulating under diverse hardware}\label{subsec:hardware_heterogeneity} 
We also consider \emph{how diverse the hardware used} in FL simulations can be and \emph{how this diversity impacts client training}.
Any FL task is defined as a distribution of isolated single-client ML tasks that are iteratively scheduled and aggregated every federated round.
Since individual client workloads are smaller than traditional ML tasks and highly parallelizable, they can be trained by many GPU types with diverse technical specifications.
Thus, for experiments involving $10^3$ to $10^4$ clients per round to be feasible, researchers and practitioners may need to leverage diverse GPUs across various servers. Furthermore, as GPU types continue to diverge in their suitability to different ML pipelines, e.g., the specialization of GPUs with very large VRAM to extremely large models, researchers and practitioners will likely accumulate GPUs of different types, generations and degrees of wear and tear.
Heterogeneous hardware environments are increasingly popular and represent a challenging setting for any general system optimization tool.

Such heterogeneity is due to several factors, such as the diversity in size of the GPUs' VRAM.
Since each client occupies limited memory, concurrently running multiple clients is affordable and beneficial to training time.
When simulating real experiments, other factors, such as system load and CPU scheduling, can affect the optimal concurrency.
This results in the need to automatically identify the proper number of workers to maximize throughput, as shown in \cref{fig:concurrency_optimum}.  
% Thus, a simulation engine must treat GPUs with larger VRAM differently from their smaller counterparts.
While the optimal strategy for some GPUs might be to pack as many clients as their VRAM allows, it might harm training throughput for GPUs with larger VRAM and lower CPU cores to GPU ratio.
Training a client roughly depends on \emph{data loading} and \emph{actual training}.
\emph{Data loading} and preprocessing generally happen on the CPU\@. The \emph{actual client training} usually happens inside the GPU, and its time is approximately determined by the average speed at which the GPU can process batches.
However, other factors contribute, such as OS scheduling, memory bandwidth, implementation efficiency, and more.
More generally, an efficient simulation engine must take advantage of the different computing systems available, irrespective of their difference in throughput.

\begin{figure}[t]
    \centering
    \noindent\includegraphics[width=0.95\columnwidth]{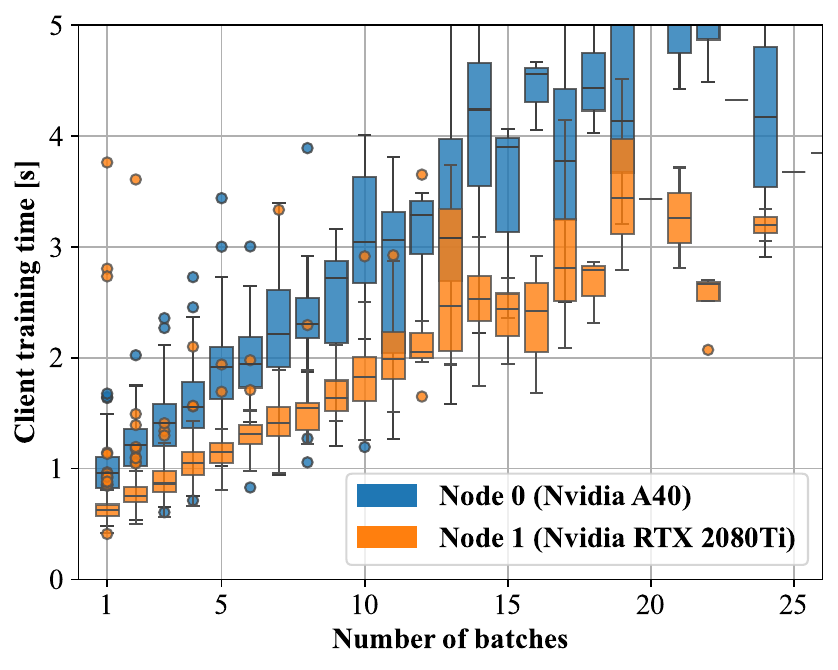}
    % \captionsetup{font=small,labelfont=bf}
    \caption{Training times for two GPUs, \textit{Nvidia A40} and \textit{Nvidia RTX 2080}, running on clients with different dataset sizes. Since the GPUs run the same clients, their diversity is reflected by the different distribution of training times with the same number of batches trained, the different trends they produce as the number of batches increases, and the pattern of the fluctuations when the training time starts to plateau.}
    \label{fig:hardware_heterogeneity}
    \Description[Figure 5]{Training times for two GPUs, \textit{Nvidia A40} and \textit{Nvidia RTX 2080}, running on clients with dataset sizes. Since the GPUs run the same clients, their diversity is reflected by the different distribution of training times with the same number of batches trained, the different trends they produce as the number of batches increases, and the pattern of the fluctuations when the training time starts to plateau.}
% \vspace{-0.4cm}
\end{figure}

\Cref{fig:hardware_heterogeneity} demonstrates that clients with the same dataset size show intra-GPU and inter-GPU variability due to the aforementioned factors.
Because of the difference between the computational systems, the pattern of the client training time differs between \textit{Node 0} and \textit{Node 1} despite training the same clients in the same order that have been sampled randomly from the entire population in advance.
Interestingly, the dispersion of groups of clients with the same number of batches in the same GPU varies depending on their size.
We also observe that, as the size and training time increase, the trend becomes unstable when it starts to plateau at around 15 batches.
This happens because of the aforementioned factors and because the client population is unbalanced, producing fewer data points for a given number of batches.

Even though the simulation engine knows the technical specifications of the computing systems used, it is very complex to use such information to proxy the training time of the simulated clients.
In particular, when leveraging concurrent workloads on the same GPU device, the OS may produce unpredictable scheduling patterns that slow down single-client training loads. By considering the variance of client training times in its learning-based placement (see \cref{sec:client_placement_models}) \pollen implicitly mitigates possible overheads of these effects.

\subsection{Communication optimization}\label{subsec:comm}
Every real FL round involves downstream communication of global model parameters to sampled clients and upstream communication of model updates from clients to the server.
Communication patterns represent another component of federated learning systems a simulation engine can optimize.
First, since the same cluster node may train multiple clients in the same round from the global model version, sending only one copy of the model to each node is possible.
Second, the server can send a list of client IDs to each node instead of sending, in many communication steps, one client ID at a time.
Third, when the aggregation algorithm allows it, the nodes can partially combine~(\cref{sec:part_agg}) the model updates of trained clients into a single model update to be communicated to the server.
\pollen is meant to effectively utilize heterogeneous resources, typical for researchers and practitioners. Thus, it optimizes for a communication topology where GPUs inside nodes are well-connected without requiring extremely fast (e.g., InfiniBand) communication across nodes. We adopt Ring-AllReduce~\citep{Horovod} to partially aggregate parameters at the node level while using a standard parameter-server approach to combine node results as soon as they become available asynchronously.
% \ls{Add table describing asymptotic complexity of our communication pattern VS other frameworks VS real FL}

\begin{figure*}[ht]
      \begin{subfigure}[t]{0.40 \textwidth}
            \caption{Federated simulation for current frameworks.}
            \label{alg:simulation_current_frameworks}
            \small
    
            \begin{algorithmic}[1]
                  \begin{gen} \STATE{Server samples a subset client IDs $C$.} \end{gen}
                  \begin{pull}
                        \STATE{Server forms $C$ into a synchronized queue $S$.}
                        \FOR{$t = 1,..., T$}
                        \WHILE{$S \neq \emptyset$}
                        \FOR{\textbf{each} worker $w \in W$}
                        \STATE{$k = \mathrm{pop}(S)$}
                        \STATE{$\theta_k^t = \mathrm{copy}(\theta^t) $}
                        \STATE{$\theta_k^{t+1} \leftarrow \mathrm{Train}_w(\theta_k^t, D_k)$}
                        \STATE{Send $\theta_k^{t+1}$ to server.}
                        \ENDFOR
                        \ENDWHILE
                        \STATE{Server aggregates trained local models to get $\theta^{t+1}$}
                        \ENDFOR
                  \end{pull}
            \end{algorithmic}

      \end{subfigure}
      \hfil
      \begin{subfigure}[t]{0.50 \textwidth}
            \caption{Federated simulation for \pollen.}
            \label{alg:simulation_pollen}
            \small
            \begin{algorithmic}[1]
                  \begin{gen}      \STATE{Server samples a subset client IDs $C$.} \end{gen}
                  \begin{push}
                        \FOR{$t = 1,..., T$}
                        \STATE{Server assigns a list of clients to train for each worker $w \in W$.}
                        \FOR{\textbf{each} node $q \in Q$}
                        \STATE{ $\theta_{q}^t = \mathrm{copy}(\theta^t)$ }
                        %\ENDFOR
                        \FOR{\textbf{each} worker $w$ inside node $q$}
                        \STATE{$\theta^{t,w} = \mathrm{copy}(\theta^t_q)$}
                        \FOR{\textbf{each} assigned client $k$ in the worker $w$}
                        %\STATE{Train a new model copy on client data.}
                        \STATE{$\theta_k \leftarrow \mathrm{Train}_w(\theta^{t,w}, D_k)$}
                        %\STATE{In-place aggregate model copy.}
                        
                        %\STATE{Node partially aggregates worker models.}
                        \STATE{$N^w_{k}=N^w_{k-1}+n_{k}$}
                        \STATE{$\theta^{t,w} \leftarrow \frac{\theta^{t,w}\times N_{k-1}+\theta_{k}\times n_{k}}{N^w_{k}}$}
                        \ENDFOR
                        \ENDFOR
                        \STATE{$q$ sends $\mathrm{RingAllReduce}(\theta^{t,w})$ to server.}
                        \ENDFOR
                        \STATE{Server aggregates partially aggregated models to get $\theta^{t+1}$}
                        \ENDFOR
                  \end{push}
            \end{algorithmic}
      \end{subfigure}
      \caption{The execution of one training round $t$ for current frameworks~(\cref{alg:simulation_current_frameworks}), and for \pollen~(\cref{alg:simulation_pollen}). We let $W$ be the set of workers; $Q$ be the set of nodes; $T$ be the total number of communication rounds; $\theta^t$ be the model parameters at round $t$; $D_k$ be the local dataset of client $k$ and $n_k = |D_k|$ be the number of sample size; Train be the local training process on the client.}
    \Description[Figure 6]{The execution of one training round $t$ for current frameworks~(\cref{alg:simulation_current_frameworks}), and for \pollen~(\cref{alg:simulation_pollen}). We let $W$ be the set of workers; $Q$ be the set of nodes; $T$ be the total number of communication rounds; $\theta^t$ be the model parameters at round $t$; $D_k$ be the local dataset of client $k$ and $n_k = |D_k|$ be the number of sample size; Train be the local training process on the client.}
% \vspace{-0.2cm}
\end{figure*}

%\subsection{Frameworks for Simulating FL}\label{subsec:simulation_engines}
%To investigate FL simulations, we consider each client operating an ML pipeline on its local data as a job.
%This job includes maintaining a copy of the model, pre-processing the data, and the training instructions.

%Given this framing, 

\subsection{Limitations of Current Systems}\label{subsec:limitations}

Simulation engines of current ad-hoc FL frameworks operate on a client-server paradigm. Workers are responsible for training clients and sending trained models to the server, which aggregates client results before the next round. The execution of a round on this system is summarised in \cref{alg:simulation_current_frameworks}.

Most of these frameworks (e.g., \flute, \fedscale, and \flower)
use a communication-heavy pull-based queuing system, where the server orchestrates FL simulations by serving clients to worker processes on GPUs. Pull-based queuing systems have two major limitations. Critically, workers cannot choose which client to train, restricting their load-balancing capabilities required to mitigate the effects of naturally partitioned datasets (\Cref{fig:client_heterogeneity}). 
Pulling a single client at a time may also add significant communication overhead, especially when clients' datasets are small and training times are comparable to fetching jobs.
One exception to this design is the recently proposed \parrot~\citep{Parrot}, which reduces communication by directly pushing a list of clients to GPUs.

Each framework also suffers from specific limitations.
\flute is constrained to using a single worker per GPU, which generally prevents it from saturating VRAM.
\fedscale loads the entire dataset on each worker, making the handling of large datasets like Reddit~\citep{Leaf} ($\backsim$\num{60}GB stored in RAM) by modest GPUs challenging.
\flower depends on Ray~\citep{Ray} as a simulation engine, which serializes scheduled jobs to disk, increasing latency.
Both \fedscale and \flower require running manual testing of hardware configurations to assess the concurrency level that the resource can afford.
In addition, the \flower's simulator does not allow setting different levels of concurrency for different GPU types, thus forcing the less capable one to be the reference.
Finally, \parrot cannot account for VRAM when balancing clients across GPUs due to using one worker per GPU\@.
%\ls{Take inspiration from PFL's paper about this. We may want to extend this section with more details and potentially move it to the appendix. Discussing some comparisons here might be useful to make a clear point about why comparing against one particular framework is like an atomic experiment for one of our specific components.}

Apple recently\footref{foot:pfl_release} open-sourced a research framework for Private Federated Learning (\pfl)~\cite{pfl_paper}. Unlike the other frameworks, it uses a design based entirely on Ring AllReduce, which can accommodate multiple workers per GPU.
Unlike \pollen, it treats all workers as members of a flat topology. Thus, it does not consider the higher intra-machine connectivity available in most clusters and data centers, especially when lacking InfiniBand, and can thus suffer when the bandwidth between the slowest two workers in the ring is low. Furthermore, it is unable to distinguish between heterogeneous GPUs and, thus, to effectively balance the workload across workers, unlike the model-based placement of \pollen. Finally, it does not offer a means of estimating the optimal number of workers that a GPU should have, which is done automatically in \pollen by the concurrency estimator. Given the complex requirements for running Ring AllReduce over an entire cluster rather than within a single machine as \pollen does, the lack of a concurrency estimator makes the practical tuning of \pfl very challenging for practitioners.

\section{\pollen Design}\label{sec:sys_design}

% To address the limitations of previous work and to meet the needs of researchers, we design \pollen to fulfill a series of complementary goals.
We describe in this section the proposed system \pollen that leverages the observations discussed in \cref{sec:sim_fl} and overcomes the limitations of the currently available systems \cref{subsec:limitations}.
\pollen fulfills a series of complementary goals to make federated learning simulations efficient and meet the needs of researchers.
The first is \textbf{maximizing experimental throughput}, which is necessary to support large-scale simulations and quick prototyping.
The second is \textbf{efficiency} in resource consumption, allowing to benefit from the \textbf{available resources}, be they \textbf{heterogeneous, considerable, or limited}.
The third is \textbf{flexibility}, which refers to the ability to run any \textbf{user-defined} ML task on any hardware with no changes to the \pollen source code.
Finally, for wide adoption to be possible, the usage of \pollen must be as \textbf{user-friendly} as possible by eliminating all hardware-related manual configuration and integrating seamlessly into the popular \flower ecosystem. 

To achieve these goals, \pollen maximizes the benefits of push-based communication by accounting for load and hardware diversity through the push-based placement system presented in this section.
Our systems' features are designed to optimize FL simulations, with particular care put into large-scale ones; they include:

\begin{enumerate}
    % \vspace{-0.1cm}
    \item \textbf{Minimal server-GPU communication}, enabled by the push-based algorithm in \cref{alg:simulation_pollen} and supported by partial aggregation. This reduces data transfer and enables \pollen to scale across multiple nodes effortlessly. 
    \item \textbf{Effective utilization} of the concurrency of each GPU, with automatic concurrency estimation. 
    \item A \textbf{load-balancing method} between GPUs using a robust concurrency-aware machine learning model for estimating client training time on a GPU. This minimizes GPU idle times to improve execution time.
% \vspace{-0.1cm}
\end{enumerate}

\begin{figure}[ht]
    \centering
    \noindent\includegraphics[width=0.93\columnwidth]{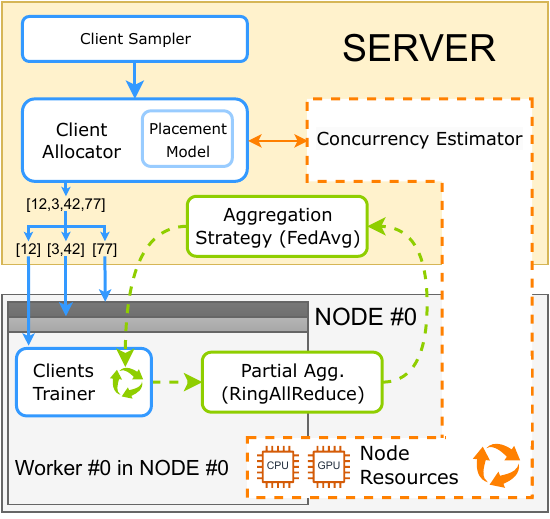}
    % \captionsetup{font=small,labelfont=bf}
    \caption{The main components of
        \pollen
        for the server, node, and worker. The colors distinguish between components of clients~(blue), models~(green), and hardware~(orange).}
    \label{fig:Fertilizer_diagram}
    \Description[Figure 7]{The main components of
        \emph{Pollen}
        for the server, node, and worker. The colors distinguish between components of clients~(blue), models~(green), and hardware~(orange).}
% \vspace{-0.5cm}
\end{figure}

\Cref{fig:Fertilizer_diagram} offers an overview \pollen's components.
The complexities of the client placement model are deferred to \cref{sec:client_placement_models}.  We emphasize that although we focus on training as the more time-intensive task, an independent pipeline built on the same principles exists for federated evaluation.

\subsection{Push-Based Placement}\label{sec:push_part}

Following the investigation in \cref{sec:sim_fl}, we propose an intelligent placement system that addresses the limitations reported in \cref{subsec:limitations} and the difficulties from \cref{subsec:client_heterogeneity,subsec:hardware_heterogeneity}.
In our approach, instead of having workers \emph{request} clients from the server, the system performs a \emph{push-based} one-shot allocation by partitioning client workloads across GPUs.
%\ls{Add diagram that describes the different communication patterns compared to pull-based systems.}
Since each GPU is hosted in a node, the server sends one copy of the global model to each node, followed by the list of client IDs assigned to each GPU on that node.
This allows us to reduce the number of server-node communication steps and to assign client sets to appropriate GPUs\@.
%\ls{Add complexity/communication analysis figures here.}
The construction and assignment of the abovementioned client ID list are deferred to the \emph{placement model}~(\cref{sec:client_placement_models}).

We \textbf{emphasize that \emph{placement}} acts on the underlying simulation layer of the framework and does not affect the FL algorithms.
It is \textbf{independent of \emph{client selection}} and executed \emph{after} client sampling for the round.
Thus, \pollen can use different sampling methods such as FedCS~\cite{FedCS}, Power-of-Choice~\cite{Power-of-Choice}, and DivFL~\cite{DivFL}.

\subsection{Concurrency Estimator}\label{sec:conc_est}

Simulations will benefit from using multiple worker processes sharing the same GPU for small individual training loads representative of cross-device training. However, dispatching too many concurrent processes leads to thrashing and an increase in overall training speed \Cref{fig:concurrency_optimum}.

We define a novel \emph{concurrency estimator} to automatically determine the concurrency level a GPU can support \emph{for a given task} without the need for manual tuning. This information is used when creating workers and by the \emph{placement model} to balance the workload between GPUs. All data collection happens at the node level, requiring no synchronization between nodes.
At the beginning of the simulation, the \emph{concurrency estimator} receives information on the available hardware of all training nodes, e.g.,~the number of CPU cores, the number and types of GPUs, and the level of concurrency they can achieve.

The \emph{maximum} concurrency level is estimated for each GPU by training one client and collecting statistics such as VRAM allocation and GPU utilization. From this point, we automatically estimate the \emph{optimal} concurrency level by linearly increasing the number of workers across rounds until the execution speed on that GPU stops improving or the maximum number of workers for the given GPU type is reached. This process happens independently and concurrently for each GPU across rounds to optimize for efficiency. To estimate the effectiveness of a given number of workers on a given GPU, we train with a concurrency level for $T_c$ rounds where $T_c$ is a user-specified hyperparameter, in the worst-case scenario, it takes the concurrency estimator $\mathcal{O}(T_c W_{\mathrm{max}})$ rounds to converge where $W_{\mathrm{max}}$ is the maximum number of total workers that the GPU type with the greatest VRAM can hold.
Thus, unlike other frameworks using multiple processes per GPU, the user does not have to constrain or manually tune the number of processes the simulation can run concurrently.

\subsection{Highly-Scalable Partial Aggregation }\label{sec:part_agg}
When using associative aggregation strategies, such as FedAvg~\cite{OgFedAvg}, we decrease communication costs and aggregation duration by partially aggregating client results within workers and nodes before sending partial results for full aggregation on the server.

We let $w$ represent the worker and $k$ represent the client, then $\theta^w_k$ be the partially aggregated model of worker $w$ after training the $k$-th client.
We also let $N^{w}_{k}$ be the total number of processed data samples after training the $k$-th client, as seen in \cref{eq:partial_agg}.

% \vspace{-0.5cm}
\begin{align}\label{eq:partial_agg}
    \theta^w_{k+1} & =\frac{\theta^w_{k}\times N_{k}+\theta_{k+1}\times n_{k+1}}{N^w_{k+1}} \\
    N^w_{k+1}        & =N^w_{k}+n_{k+1}
\end{align}
% \vspace{-0.5cm}

After all workers have finished training, the node computes a \emph{partial update $\theta^w$} to be sent to the server by aggregating the worker models using Ring AllReduce. In a scenario where each GPU only hosts one worker and intra-machine GPUs are well connected, Ring AllReduce is bandwidth-optimal~\citep{Horovod} as it scales linearly in the number of parameters and only depends on the lowest bandwidth connection between GPUs rather than the number of GPUs. Despite these advantages, Ring AllReduce cannot be applied for federated simulations without consideration for the practical execution environments researchers or practitioners may have, as done by other works~\citep{pfl_paper}. Since federated learning only requires communicating once per round rather than once per batch, it makes little sense to impose the exact stringent bandwidth requirements upon the hardware used to simulate it, as using more numerous but poorly connected GPUs will still increase efficiency. As such, \pollen only executes Ring AllReduce within a given machine as to allow it to use the greatest breadth of hardware accessible. 

We still only communicate once from the server to the workers for the few FL algorithms that are not associative, such as Robust Aggregation~\citep{RobustAggregation}.
However, we return a packet of all client models from the workers to the nodes and then the server. 
%\ls{Expand on the use of ring all reduce.}
% \ls{Show numbers related to concurrency and ring all reduce.}
% \ls{Discuss numerical instabilities}
% \ls{complexity and runtime analysis. How much does it take? How does this depend on the number of workers? Involve horovod paper.}

\section{Client Placement Model}\label{sec:client_placement_models}

Research on large-scale~\citep{LargeCohorts,FLINT} FL often relies on simulating FL settings with $[10^3, 10^4]$ clients per round, corresponding to populations of millions over relatively constrained hardware settings for thousands of rounds.
We expect any one-shot placement method~(\cref{alg:simulation_pollen}) to outperform the communication design of other works and improve simulation times.
However, we argue that reducing the experiment latency as much as possible while maximally exploiting hardware resources is necessary for genuine large-scale research and will become more critical as client populations grow. 
In this section, we analyze the complexities of \emph{intelligent client placement} as the most promising means of further improving simulation time and justify our choice of a concurrency-aware placement model for \pollen.

\begin{figure}[t]
    \centering
    \noindent\includegraphics[width=\columnwidth]{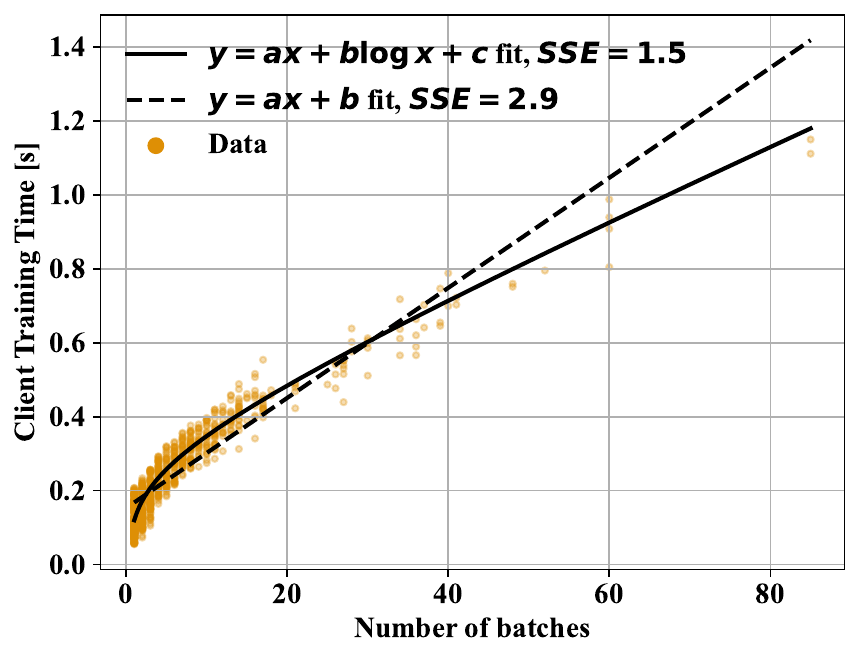}
    % \captionsetup{font=small,labelfont=bf}
    \caption{Clients' training times plotted against their number of batches. The fitting lines of the linear and proposed functions are shown alongside the fit's Summed Squared Errors (SSE).}
    \label{fig:LB_fit_function}
    \Description[Figure 8]{Clients' training times plotted against their number of batches. The fitting lines of the linear and proposed functions are shown alongside the fit's Summed Squared Errors (SSE).}
% \vspace{-0.5cm}
\end{figure}

\subsection{Placement Structure}\label{sec:EfficientClientPlacement}

While \pollen improves the communication efficiency of previous works through its \emph{push-based} client placement, the one-shot allocation used for optimizing communication prohibits runtime client reallocation between nodes.
Thus, to balance clients across GPUs and avoid stragglers, the server must predict the training time of each client on a GPU.
As shown in \cref{subsec:client_heterogeneity}, estimating training time based on only client size and systems specifications is unfeasible.
Furthermore, the estimated concurrency~(\cref{sec:conc_est}) must also be considered when balancing.

Our design allows the simulation to have only one server-node communication step per round, and the allocation to a node is done for each of its GPUs.
Thus, the placement model for \pollen must distribute a list of $N$ randomly sampled clients across GPUs in one step.
Additionally, for a model to be concurrency-aware, it must consider the number of workers of a GPU.
We achieve this by allocating clients per worker in a GPU rather than the entire GPU.
Due to the clear benefits of concurrency, we only consider concurrency-aware models in our analysis. 

For each round, a placement model receives a list of sampled client IDs together with the features of each client, and it returns a list of clients' IDs for each worker, indicating which clients to train.
To generalize across FL tasks, we rely on the number of batches a client has to predict its training time.
However, when clients can have different values for a given ML pipeline parameter, such as batch size, \pollen can be extended to use these parameters as additional features. 

To provide a clear baseline for our model, we use two methods that do not require knowledge about training times.

\noindent\textbf{Na\"{\i}ve Round-Robin (RR) Placement:}
This simple placement method splits the list of clients into $w$ uniformly populated lists, $w$ being the number of total workers.
Possible remainders are distributed across the first workers.
This client placement method is simple and robust for homogeneous client loads and hardware settings.
However, diversity in terms of hardware~(\cref{subsec:hardware_heterogeneity}) and client loads~(\cref{subsec:client_heterogeneity}) results in an unbalanced allocation across GPUs.
For example, if one GPU is twice as fast as another with the same amount of VRAM, Na\"{\i}ve Round-Robin (RR) would allocate them similar loads, causing idle time. 

\noindent\textbf{Batches-Based (BB) Placement:}
The natural piece of information needed to improve upon RR is the number of batches $x$ held by a given client as a proxy for the training time of the client. Thus, this second baseline method allocates work across workers by balancing the number of batches their clients hold.  However, Batches-Based placement understands neither how training time scales with batch count nor the speed difference between GPUs.

\subsection{Learning-Based (LB) Placement}

As indicated for \emph{Batches-Based} placement, due to the intra and inter-GPU training time variability shown in \cref{fig:hardware_heterogeneity}, the number of batches $x$ alone is an insufficient proxy for client load.
Thus, we propose a learning-based model for \pollen, which predicts the training time for an individual GPU\@.
A side objective of \pollen's model is to execute the fitting procedure quickly, as this is repeated with an increasing amount of data throughout the entire experiment.

The placement model of each GPU receives a dataset of tuples $(client~training~time,~x)$ for which it fits the curve $f(x)$ of a given number of batches $x$, as shown in \cref{eq:lb_model}.

% \vspace{-0.3cm}
\begin{equation}\label{eq:lb_model}
    f(x)=ax+b\log\left(cx\right)+d
\end{equation}

The first two FL rounds will use the \emph{Na\"{\i}ve Round-Robin} placement method to collect unbiased training time data from all available workers.
From the third round onwards, \pollen switches to using the parameters derived from fitting \cref{eq:lb_model} on previous data.
Because the fitting happens while clients are training, it uses data generated up to round $t-2$ to fit the curve for round $t$.
After every round, \pollen appends new data for the fitting procedure. 
We chose \cref{eq:lb_model} to match the skewed empirical distribution of training time shown in \cref{fig:LB_fit_function} and will further justify this choice in \cref{sec:client_placement_models:adaptive_log_linear}.

Using the predicted training time for the squared batch size, the \emph{Learning-Based (LB)} placement method then sorts the workers by GPU type, from the fastest to the slowest.
Finally, the LB method balances client load across workers.
It achieves this by ordering clients by $x$ from largest to smallest and assigning them to workers based on worker order, e.g., at the start the largest client will go to the fastest worker.
After each assignment, the workers are re-sorted based on the sum of the loads assigned to them. 
While this placement method is designed to be highly robust, we enhance it to allow short-term adaptivity to recent data.

\subsubsection{Robust and Adaptive Modelling}\label{sec:client_placement_models:adaptive_log_linear}

The robustness of Learning-Based placement relies on the fitting function and the data used for the prediction.
The number of data points generally determines the robustness and the duration of the curve fitting.
Concerning data, \pollen keeps all data from previous rounds unless a time window for deleting old data is specified.
Because we fit the curve offline, this does not increase simulation time.
Thus, the overhead of LB over RR is near zero, as it includes only the prediction and sorting times.
Concerning the fitting function, the log-linear curve shown in \cref{eq:lb_model} was chosen for its mathematical properties based on several factors. 

First, since small clients have greater training time variance, many may take longer than their slightly bigger counterparts.
This behavior may enforce negative slopes in the fitted curve, especially for polynomial functions.
The log-linear curve ensures that the fitted function never predicts negative values despite the vast cloud of data points produced by small clients, as observed in \cref{fig:LB_fit_function}.
\emph{As such, our function allows us to avoid ever predicting a negative time for a client at the cost of slightly overestimating the duration of small clients.} 

Second, the relationship between $x$ and training time is complex, as described in \cref{sec:sim_fl}.
The logarithmic term makes the function more robust, while the linear term ensures that the larger clients are predicted to take longer.
Notably, the log-linear curve can always fit linear behavior, while a linear function cannot handle situations where the underlying relationship is logarithmic.
This empirically derived choice is one of the critical differences between \pollen and \parrot.

To account for clients who deviate from the expected trend and for time-dependent effects on system performance~(e.g., node workload shifts), \pollen uses adaptive error correction.
Specifically, we correct the prediction $f(x)$ to $g(x)$ using \cref{eq:err_corr}, where the correction term inside the parentheses is the average training time for $x$ observed in recent data.
For our experiments, we use data from the most recent round.

% \vspace{-0.2cm}
\begin{equation}\label{eq:err_corr}
    g(x)= \tfrac{1}{2} \left(f(x) + \tfrac{1}{r}\textstyle\sum_{t-2-r}^{t-2}\overline{X_i} \right)
\end{equation}
% \vspace{-0.4cm}

\section{Experimental Design}\label{sec:experimental_design}
This section describes our experimental setup.
The codebase for reproducing the experiments will be made publicly available.
Although absolute numbers produced by the experiments will strongly depend on hardware specifications, we will report meaningful metrics for comparison across hardware settings.
Notably, we isolated framework simulations to prevent interference effects, and we only compared results between frameworks for the same hardware setting.
To be fair towards frameworks that use multiple processes per GPU without being able to estimate the proper amount of concurrency, we determine the optimal number of workers to allocate to each GPU prior to running experiments for all frameworks. To avoid being at a disadvantage due to having to estimate the optimal concurrency, we run the concurrency estimator prior to the full experiment and then use the number it outputs from the start.
One notable exception is the \emph{Masked Language Modeling} task, described below, in the \fedscale framework.
The limitations of \fedscale's data loading design (\cref{subsec:limitations}) did not allow more than one process due to RAM constraints.

\begin{figure*}[t]
\begin{minipage}[t]{0.48\textwidth}
\noindent\includegraphics[height=0.7\columnwidth]{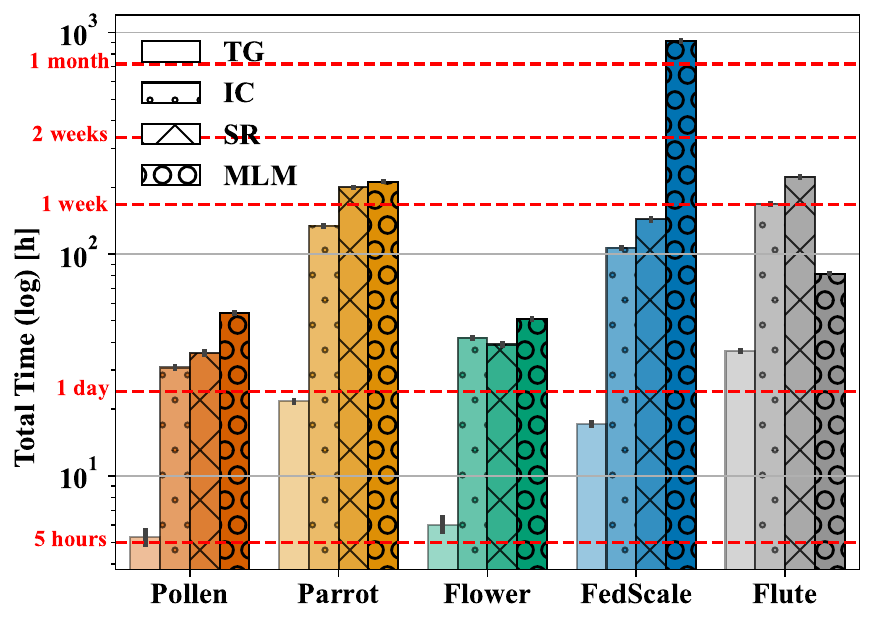}
    % \captionsetup{font=small,labelfont=bf}
    \caption{Single-node framework comparisons for \emph{medium-scale} settings. \pollen reduces communication and exploits concurrency to speed up simulations even on homogeneous hardware.}
    \label{fig:Framework_Comparison}
    \Description[Figure 9]{Single-node framework comparisons for \emph{medium-scale} settings. \pollen reduces communication and exploits concurrency to speed up simulations even on homogeneous hardware.}
\end{minipage}
\hfill
\begin{minipage}[t]{0.48\textwidth}
\noindent\includegraphics[height=0.7\columnwidth]{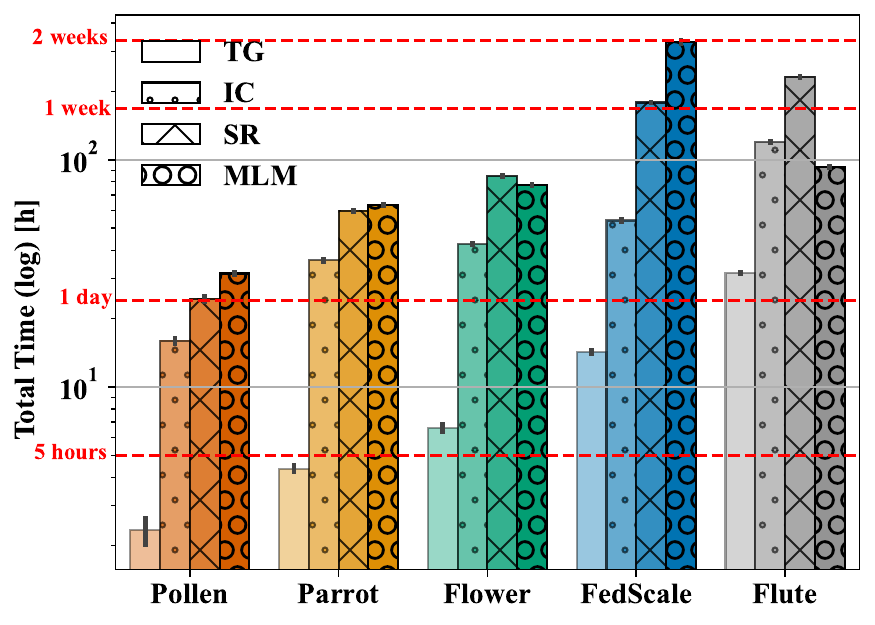}
    % \captionsetup{font=small,labelfont=bf}
    \caption{Multi-node framework comparisons for \emph{medium-scale} settings with heterogeneous hardware. \pollen outperforms others via its machine-learning model for client placement.}
    \label{fig:Multinode_Framework_Comparison}
    \Description[Figure 10]{Multi-node framework comparisons for \emph{medium-scale} settings with heterogeneous hardware. \pollen outperforms others via its machine-learning model for client placement.}
\end{minipage}
% \vspace{-0.3cm}
\end{figure*}

\begin{table}[ht]
    \centering
    \caption{The \emph{number of clients per round} used for the scalability experiments. The header codes identify the tasks.}
    \begin{tabular}{rcccc}
        \toprule
        \textbf{Scale} & \textbf{TG} & \textbf{IC} & \textbf{SR} & \textbf{MLM} \\
        \hline
        \emph{Medium Scale} & 100 & 100 & 100 & 100 \\
        \emph{Large Scale} & 1000 & 1000 & 1000 & 1000 \\
        \emph{Very Large Scale} & 10000 & 10000 & 2000 & - \\
        \bottomrule
    \end{tabular}
    % \vskip +5pt
    \label{tab:scales_exp}
% \vspace{-0.7cm}
\end{table}

\subsection{Federated Learning Tasks}
We use four representative FL tasks throughout this work.
Each task comes with its model architecture and its naturally federated dataset.
We exclude clients without enough samples to fill a single batch across all tasks.
Every experiment uses Federated Averaging with one local epoch.

\noindent\textbf{Image Classification (IC):} The goal of this task is to collaboratively train a ShuffleNetV2~\citep{shufflenet} network to classify images amongst \num{596} classes.
We use a federated version of the original Open Image~\citep{OpenImages} dataset as implemented by \fedscale, which contains \num{1,6e6} images partitioned across \num{13771} clients.

\noindent\textbf{Speech Recognition (SR):} This task uses the Google Speech Commands dataset~\cite{Google_Speech} to collaboratively train a ReseNet-34~\cite{resnet} network to classify audio samples amongst a set of $35$ pre-defined spoken words.
The dataset contains a collection of \num{157e3} one-second-long clips naturally partitioned according to their \num{2168} speakers.

\noindent\textbf{Text Generation (TG):} We use the Shakespeare dataset, as implemented in TensorFlow Federated~\citep{tensorflowfederated}, to train a two-cell LSTM-based language model as defined in~\citep{Leaf}.
The dataset comprises sentences extracted from \textit{The Complete Works of William Shakespeare} and grouped into \num{648} fictional characters.

\noindent\textbf{Masked Language Modelling (MLM):} This task uses the Reddit dataset~\citep{Leaf}, a collection of comments posted on Reddit.
It represents \num{1,6e6} clients for more than \num{56e6} samples.
We follow the experimental configuration proposed in the \fedscale's example, using the \emph{RoBERTa}~\citep{RoBERTa} model with the \emph{Albert-V2-tokenizer} \citep{AlbertToken}.

\subsection{Hardware Configurations}\label{subsec:hard_conf}
Using two representative GPU types, we consider two hardware configurations that reflect common research clusters, namely \emph{single-node} and \emph{multi-node}.
Experiments were run on the Slurm~\citep{SLURM} scheduler to respect common environmental conditions of computing clusters.

\noindent\textbf{Single-node:} Our \emph
{single-node} experiments were run on \emph{node 0} containing Nvidia A40 GPUs and an Intel (R) Xeon (R) Gold 6152 with \num{88} cores.
One A40, due to the scheduler limitations, is paired with \num{11} CPU cores out of \num{88}.
Our experiments use a \emph{homogeneous} setup with \emph{a single GPU} on this node, i.e.~one Nvidia A40 with 11 CPU cores.

\noindent\textbf{Multi-node:} Our \emph{multi-node} experiments were run on a combination of the aforementioned \emph{node 0} containing A40s and \emph{node 1}.
\emph{Node 1} contains Nvidia RTX 2080 Ti GPUs and an Intel(R) Xeon(R) CPU E5-2680 v4 containing $56$ cores.
Each GPU in \emph{Node 1} is paired with \num{8} CPU cores out of \num{56}.
In this \emph{diverse} setting, we use one Nvidia A40 from \emph{node 0} paired with \num{3} Nvidia 2080 Ti to balance the VRAM across the two GPU types.
After the workers have completed their partial aggregation on the \emph{CPU} of their node, final aggregation steps happen on \emph{node 0} on the \emph{CPU}.

\subsection{Framework Benchmark}
This set of experiments aims to evaluate the impact of abandoning previous pull-based queuing systems in favor of our push-based solution.
We compared \pollen against the other frameworks by measuring the time different simulators take to finish \emph{medium scale} experiments. Since it does not have an openly-available implementation, we implement the design of \citet{Parrot} using the same backend as \pollen.
The four tasks are executed under the two hardware settings mentioned above.
The results of these experiments are presented and discussed in \cref{subsec:framework_singlenode,subsec:framework_multinode}.

\subsection{Cohort size scalability}
Our main contribution in this paper is to make efficient large-scale simulations feasible at the scale of real-world applications.
To evaluate this goal, we aim to measure how \pollen performs compared with the other frameworks as we \emph{increase the federated population size} and the \emph{number of clients per round} accordingly.
Following \citet{ScaleAndSystemDesign}, we always sample $0.1\%$ of the simulated population, e.g.,~we sample $10^4$ clients per round from a population of $10^7$.
When the dataset has an insufficient population size, we sample with replacement.

We chose the scales for the tasks as reported in \cref{tab:scales_exp}.
We note that for \emph{very large scale} experiment, we were forced to decrease the scale for \emph{Speech Recognition} and to abandon the experiment for \emph{Masked Language Modeling}.
This is because exciting other frameworks would require unreasonable times at those scales, making such a comparison meaningless. However, we put the results from \pollen in the appendix
The results for all frameworks are in \cref{subsec:scalability_eval}.

\subsection{Model prediction}
Finally, we evaluate the impact of the placement model's predictions on GPU utilization and idle time.
We assessed how effectively \pollen can capture information regarding which GPU devices would increase the speed of the simulation by receiving a more extensive workload.
To this aim, we measure the GPU utilization in our \textit{multi-node} configuration and idle time compared \pollen against the RR baseline from \cref{sec:EfficientClientPlacement}.
We run this experiment for each task, spanning the time frame of ten rounds.
This comparison is discussed in \cref{sec:idle_time_evaluation}.
\section{Evaluation}\label{sec:eval}

We show that \pollen delivers faster FL simulations than \flower~\cite{Flower}, \flute~\citep{Flute}, \fedscale~\citep{FedScale}, \parrot~\cite{Parrot}, and \pfl~\cite{pfl_paper}.
Then, we show that \pollen's benefits grow when populations increase from thousands to millions. They improve from saving researchers days at medium scales to weeks/months at large ones, making it the \emph{only} reasonable choice.
Finally, we prove that \emph{learning-based} placement reduces GPU idle time, beating the \emph{Round-Robin} and \emph{Batches-Based} baselines.

% \vspace{-0.1cm}
\begin{figure}[ht]
    \centering
    \noindent\includegraphics[height=0.7\columnwidth]{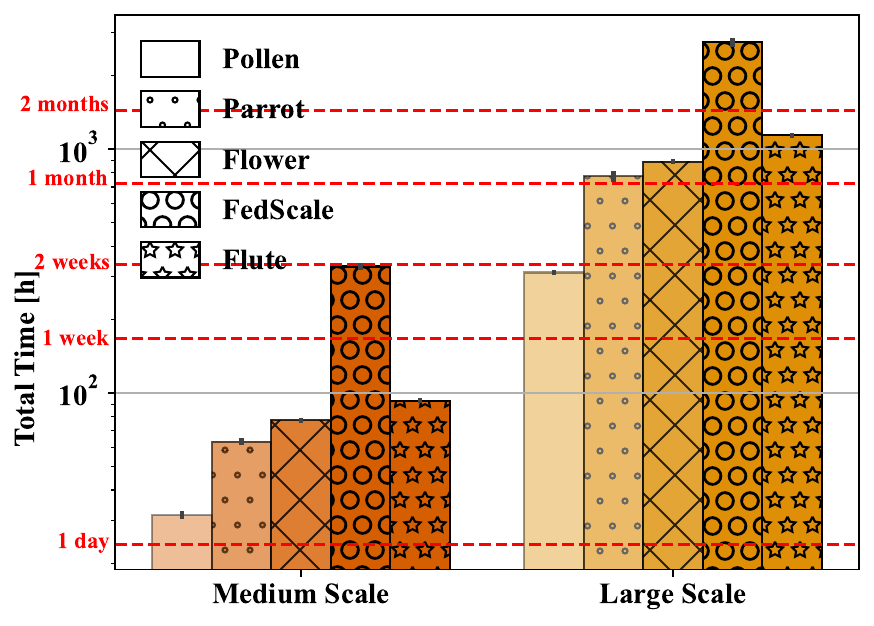}
    % \captionsetup{font=small,labelfont=bf}
    \caption{Scalability comparisons for \emph{Masked Language Modelling}. \pollen brings improvements of days/weeks for \emph{medium-scale} settings and weeks/months  for \emph{large scale}.}
    \label{fig:Scale_MLM}
    \Description[Figure 11]{Scalability comparisons for \emph{Masked Language Modelling}. \pollen brings improvements of days/weeks for \emph{medium-scale} settings and weeks/months  for \emph{large scale}.}
% \vspace{-0.5cm}
\end{figure} 

\subsection{Homogeneous Single-node Comparisons}\label{subsec:framework_singlenode}
For \emph{single-node}~(\cref{subsec:hard_conf}) execution with one Nvidia A40 GPU, \cref{fig:Framework_Comparison} indicates that across most experiments \pollen improves simulation time over previous works, even for constrained hardware. 
The simulator of its closest competitor, \flower, operates similarly to \pollen for our \emph{single-node} setting where hardware diversity is nonexistent.
Thus, for all except the \emph{Masked Language Modeling}, the FL-specific engineering that \pollen uses to exploit the small size of client jobs allows it to improve upon \flower in this setting. 

The other frameworks underperform \pollen by hours/days for the simple \emph{Text Generation} task and days/weeks for all others.
Overall, this hardware-\emph{homogeneous} \emph{single-node} setting indicates the impact of push-based communication. Furthermore, the comparisons against \flute and \parrot prove the importance of using multiple concurrent workers per GPU.
We argue that \fedscale fails to obtain a similar advantage to \pollen and \flower from concurrency due to inefficient communication and data-loading, which bottleneck the workers' throughput, as mentioned in \cref{subsec:limitations}.

\textbf{Implications:}~\pollen brings notable speedups over previous works, even for constrained hardware settings. 
Thus, it fulfills its goals of \emph{maximizing throughput} and \emph{flexibly} adapting to available resources.
Furthermore, its minimal communication and effective concurrency utilization are validated. 
Consequently, \pollen boosts the speed of prototyping and experimentation for researchers with limited resources. 

\begin{figure*}[ht]
      \begin{subfigure}[t]{0.33 \textwidth}
            \centering
            \noindent\includegraphics[height=0.7\columnwidth]{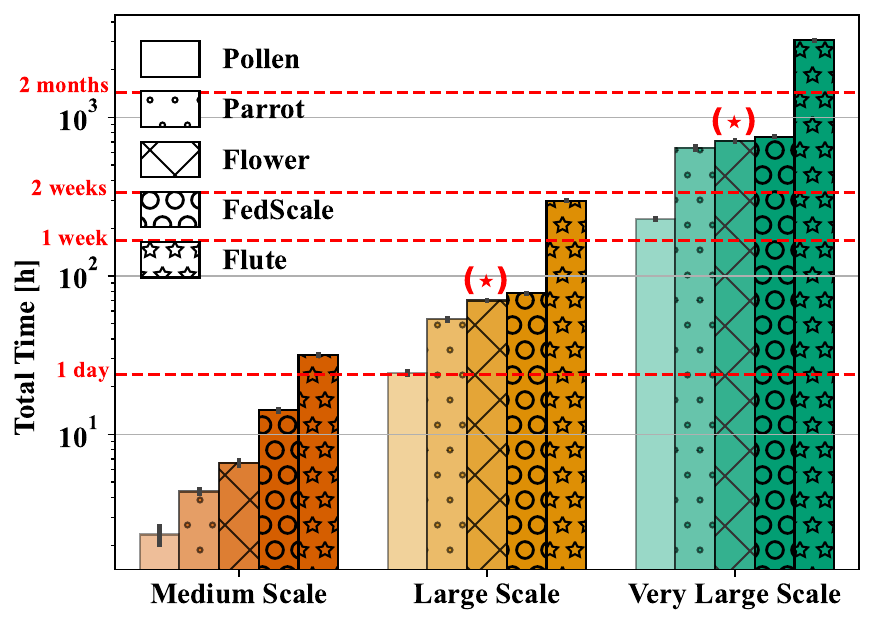}
            % \captionsetup{font=small,labelfont=bf}
      \end{subfigure}
      \hfil
      \begin{subfigure}[t]{0.33 \textwidth}
        \centering
        \noindent\includegraphics[height=0.7\columnwidth]{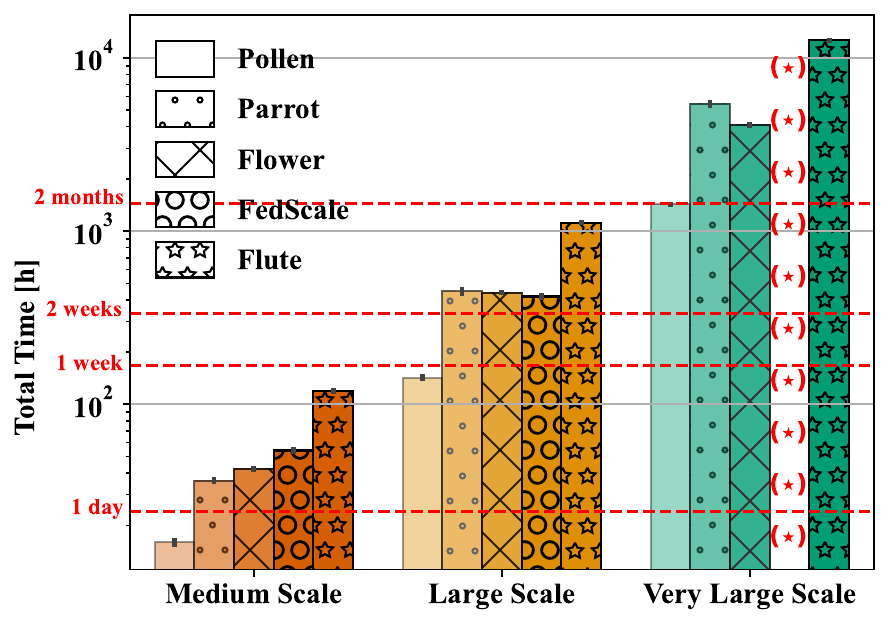}
        % \captionsetup{font=small,labelfont=bf}
      \end{subfigure}
      \hfil
      \begin{subfigure}[t]{0.33 \textwidth}
        \centering
        \noindent\includegraphics[height=0.7\columnwidth]{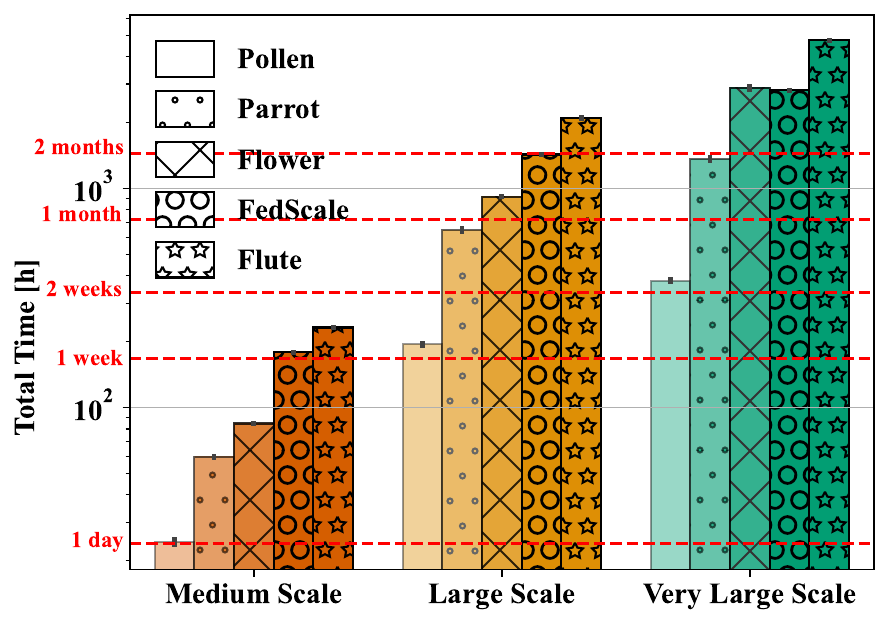}
        % \captionsetup{font=small,labelfont=bf}
      \end{subfigure}
    \caption{Scalability comparisons for \emph{Text Generation}~(left), \emph{Image Classification}~(center), and \emph{Speech Recognition}~(right). \pollen outperforms by hours~(\emph{medium-scale}), days/weeks~(\emph{large-scale}), and weeks/months~(\emph{very large scale}). Asterisks indicate training failures resulting in incomplete rounds or the impossibility of executing the experiment.}
    \label{fig:Scale_TG-IC-SR}
    \Description[Figure 12]{Scalability comparisons for \emph{Text Generation}~(left), \emph{Image Classification}~(center), and \emph{Speech Recognition}~(right). \pollen outperforms by hours~(\emph{medium-scale}), days/weeks~(\emph{large-scale}), and weeks/months~(\emph{very large scale}). Asterisks indicate training failures resulting in incomplete rounds or the impossibility of executing the experiment.}
\end{figure*}

\subsection{Diverse Hardware Multi-node Comparisons}\label{subsec:framework_multinode}
The diversity of hardware in the \emph{multi-node}~(i.e.,~2 nodes, the first equipped with 1 Nvidia A40 and the second with 3 Nvidia 2080s Ti, as specified in \cref{subsec:hard_conf}) setting magnifies the improvements in simulation time brought by \pollen in the previous one.
\Cref{fig:Multinode_Framework_Comparison} shows \pollen outperforming for all experiments, increasing its lead against competitors across most tasks.
The other frameworks continue underperforming by hours/days for \emph{Text Generation} and days/weeks for all others.

Beyond generalizing the benefits of push-based client placement and concurrency from \emph{single-node} settings, these results evidentiate the importance of \emph{client placement}.
Balancing client loads across diverse GPUs is thus paramount, given the inefficiency of inter-node client reallocation.
For example, in \emph{multi-node} settings \flower's simulator falls behind \pollen's as it is \emph{pull-based}.
Additionally, the concurrency advantage and robust log-linear model allow \pollen to continue outperforming \parrot despite its push-based communication. 

\textbf{Implications:} \pollen significantly improves execution time for \emph{multi-node} settings due to its communication efficiency and robust concurrency-aware client placement model. Furthermore, we argue that \pollen can scale effortlessly over even larger clusters because of these features.
Since \pollen did not require additional configuration to handle the \emph{multi-node} setting, we have shown it to be both \emph{flexible} and \emph{user-friendly}.
As such, we argue that \pollen is the natural choice for researchers with large distributed clusters.

\begin{table*}[ht]
    \centering
    \caption{GPU utilization [\%] and total GPU idle time [s] measured over the first ten federated rounds for every task executed in our \textit{multi-node} setting. The GPU statistics are reported separately to show that \pollen successfully recognizes that the \textit{RTX2080} GPUs are quicker at executing the tasks investigated and thus increase their utilization compared to the \textit{A40} GPU. Note that, as described in \cref{subsec:hard_conf}, the amount of VRAM is balanced, and the number of workers in the \textit{A40} GPU is approximately the same as the sum of the workers in the three \textit{RTX2080} GPUs.}
    % \begin{tabular}{T{0.14\columnwidth}cccccccc}
    \begin{tabular}{rcccccccc}
        \toprule
         \textbf{\textit{GPU}} & \multicolumn{2}{c}{\textit{Utilization A40 [\%]}} & \multicolumn{2}{c}{\textit{Utilization RTX2080s [\%]}} & \multicolumn{2}{c}{\textit{A40 idle time [s]}} & \multicolumn{2}{c}{\textit{RTX2080s idle time [s]}} \\
         \textbf{\textit{metrics}} & \textbf{\pollen} & \textbf{RR} & \textbf{\pollen} & \textbf{RR} & \textbf{\pollen} & \textbf{RR} & \textbf{\pollen} & \textbf{RR} \\
        \hline
        \textbf{SR} & \textbf{19.4 $\pm$ 0.3} & 19 $\pm$ 1 & \textbf{7.4$\pm$0.5} & 6.9 $\pm$ 0.8 & 77 $\pm$ 16 & \textbf{68 $\pm$ 17} & \textbf{93 $\pm$ 15} & 115 $\pm$ 25 \\
        \textbf{TG} & 45 $\pm$ 3 & \textbf{51 $\pm$ 2} & 16 $\pm$ 4 & \textbf{17 $\pm$ 2} & 14 $\pm$ 10 & \textbf{9 $\pm$ 8} & 53 $\pm$ 9 & \textbf{43 $\pm$ 6} \\
        \textbf{IC} & 61.7 $\pm$ 0.4 & \textbf{68.1 $\pm$ 0.8} & \textbf{30 $\pm$ 9} & 18 $\pm$ 3 & 26.4 $\pm$ 0.3 & \textbf{18 $\pm$ 4} & \textbf{71 $\pm$ 22} & 114 $\pm$ 12 \\
        \textbf{MLM} & 60.7 $\pm$ 0.5 & \textbf{66.3 $\pm$ 0.9} & \textbf{27 $\pm$ 3} &  19 $\pm$ 4 & 45 $\pm$ 12 & \textbf{27.4 $\pm$ 0.8} & \textbf{133 $\pm$ 13} & 183 $\pm$ 34 \\
        \bottomrule
    \end{tabular}
    % \vskip +5pt
    \label{tab:gpu_util_comparison}
% \vspace{-0.4cm}
\end{table*}

\subsection{Scalability Comparisons}\label{subsec:scalability_eval}
When evaluating FL frameworks on large and very large scales in \emph{multi-node}~(\cref{subsec:hard_conf}) simulations, we show that \pollen's improvements compound to bring unprecedentedly fast simulations for these crucial federated settings.

The \emph{Text Generation} task in \cref{fig:Scale_TG-IC-SR} is the quickest; however, it shows increasingly large gaps between \pollen and all other frameworks as scales increase. Thus, it proves that simulation efficiency becomes a bottleneck for large-scale research, even for simple datasets.
The other tasks show similar behaviors with longer time scales.
Thus \pollen saves months of training time for very large experiments over \emph{all} competitors.
Notably, for \emph{Speech Recognition} on the largest-scale settings, \pollen can execute in two weeks while the fastest competitor requires two months.

Other frameworks, \fedscale and the original \flower, also show occasional unreliability.
For \emph{Image Classification}, \fedscale fails to execute a single round in the very large scale setting because it cannot aggregate all the client models. 
In the case of \emph{Text Generation}, the simulator of \flower drops large numbers of clients.

\textbf{Implications:} \pollen is the only FL simulator to enable research at the scales necessary for investigating large-cohort effects.
All other systems require months to execute when weeks suffice.
Thus, \pollen has the sole potential to match the growth of FL as it extends globally.

\subsection{Placement Model Efficiency Evaluation}\label{sec:idle_time_evaluation}
The LB placement model of \pollen learns how to efficiently accommodate clients to the resources available, also discriminating which GPUs are training faster than the others.
We benchmarked in \cref{tab:gpu_util_comparison} the GPU utilization and the GPU idle time for the first \num{10} rounds of the \textit{medium scale} experiments for each task running on our \textit{multi-node} configuration.
In the limited time frame of \num{10} rounds, \pollen recognizes that the three \textit{Nvidia RTX2080} have to be preferred and tries to accommodate more clients to those GPUs to increase their utilization and reduce their idle time.

\textbf{Implications:} \pollen can push GPUs to higher utilization when they are likely to improve the speed of the experiment by training more clients per round.

\begin{table*}[t]
    \centering
    \caption{The duration of the \textbf{full aggregation} at the server using \textit{FedAvg} for all the tasks.
    % Three different values for the number of clients/models have been used, i.e., 10, 100, 1000, to show how such a duration scales.
    Two different values for the number of clients/models have been used, i.e., 10 and 100, to show how such a duration scales.
    Note that the different tasks are representative of the models' scales: TG weighs 3.28MB, IC weighs 26.45MB, MLM weighs 60.37MB, and SR weighs 85.14MB.
    For reference, the mean and standard deviation of the \textit{round duration} for \pollen and \flower are given below
    % the settings with 100 and 1000 clients.
    the setting with 100 clients per round.
    The gap between these two references is always one order of magnitude greater than the corresponding \textit{full aggregation time}, showcasing that \pollen's speed-up is not only due to \textit{partial aggregation}'s contribution.
    % Note that \flower uses full aggregation and \pollen uses partial aggregation.
    }
    \begin{tabular}{rcccccc}
    \toprule
     & \textit{Cohort Size} & \textbf{TG} [s] & \textbf{IC} [s] & \textbf{MLM} [s] & \textbf{SR} [s] \\ 
    \hline
    \textit{Full} & \textbf{10} & 0.01 $\pm$ 0.001 & 0.08 $\pm$ 0.01 & 0.54 $\pm$ 0.37 & 1.86 $\pm$ 0.89 \\
    \textit{Aggregation} &\textbf{100} & 0.08 $\pm$ 0.01 & 5.18 $\pm$ 2.17 & 1.68 $\pm$ 0.17 & 13.20 $\pm$ 4.18 \\
    \hline
    \textit{Federated} & \textbf{\pollen}-100 & 1.6844  $\pm$ 0.0006 & 11.4992 $\pm$ 0.0009 & 22.7298 $\pm$ 0.0008 & 17.593 $\pm$ 0.001 \\
    \textit{Round} & \textbf{Flower}-100 & 4.7532 $\pm$ 0.0005 & 30.5190 $\pm$ 0.0006 & 55.7132 $\pm$ 0.0004 & 60.936 $\pm$ 0.001 \\
    \bottomrule
    % \textit{Full} & \multirow{2}{*}{\textbf{1000}} & \multirow{2}{*}{2.99 $\pm$} & \multirow{2}{*}{27.81 $\pm$} & \multirow{2}{*}{74.08 $\pm$} & \multirow{2}{*}{96.09 $\pm$} \\
    % \textit{Agg.} & \\
    % \hline
    % \textit{Fed.} & \textbf{Pollen}-1000 & 17.618 $\pm$ 0.001 & 102.580 $\pm$ 0.003 & 224.667 $\pm$ 0.001 & 139.925 $\pm$ 0.006 \\
    % \textit{Round} & \textbf{Flower}-1000 & 50.3368 $\pm$ 0.001 & 317.214 $\pm$ 0.004 & 639.994 $\pm$ 0.008 & 660.891 $\pm$ 0.002 \\
    % \hline
    \end{tabular}
    % \captionsetup{font=small,labelfont=bf}
    % \vskip +5pt
    \label{tab:part_agg_benchmark}
% \vspace{-0.6cm}
\end{table*}
% \vspace{-0.5cm}
\subsection{Partial Aggregation contribution}\label{sec:part_agg_benchmark}

The \textit{highly-scalable partial aggregation} component described in \cref{sec:part_agg} enables \pollen to efficiently accommodate large cohort size in every federated round of an experiment.
In \cref{tab:part_agg_benchmark}, we benchmark the speed-up introduced by partial aggregation per single federated round by comparing the amount of time that one \textit{full aggregation} step requires for the different tasks with the federated round time of \pollen and \flower.
Notably, \flower always uses \textit{full aggregation} while \pollen uses \textit{partial aggregation}.
\flower represents the perfect baseline because, relying on \textit{Ray}\cite{Ray}, it uses a very fast in-memory communication protocol to accumulate model updates to the server whose overhead is minimal.
Also, since \pollen is built on top of \flower, they share most of the abstraction in the implementation, e.g.,~the functions extracting the model parameters.
The difference between the federated round time between \flower and \pollen is always one order of magnitude greater than the amount of time that full aggregation takes.

\textbf{Implications:} \textit{Partial aggregation} represent an essential component of \pollen even though the placement model, concurrency estimator, and the push-based design contribute most of the improvement compared to \flower, which is the closest framework.

\subsection{Comparison against \pfl}\label{sec:pfl_comparison}

We compared against \pfl using the two representative tasks, namely CIFAR-10~\cite{krizhevsky2009learning} and FLAIR~\cite{flair_paper}, in \cref{tab:pfl_comparison}.
The CIFAR-10 dataset has been artificially partitioned into \num{1000} balanced clients\footnote{When the cohort size is bigger than the federated population, it is compo-\\sed of randomly sampled clients with replacement. Notably, \pfl performs\\sampling with replacement by default irrespective of the population size or\\the cohort size.}. FLAIR is naturally partitioned with a skewed distribution, where most clients hold very few samples.
As \pollen's main objective is to allow for large-scale experimentation, we extended the baseline settings~\cite{pfl_wandb_report} to cohort sizes of different orders of magnitude.
We used our \textit{multi-node} hardware configuration for all the experiments related to the CIFAR-10 dataset. For FLAIR, because \pfl had difficulties
% running multiple workers per GPU at the larger scales
, we introduce a new hardware setting identical to our previous \emph{multi-node} one except for being equipped with only one Nvidia RTX 2080 Ti GPU. 

\textbf{Implications:} Simply utilizing highly efficient distributed training algorithms is insufficient to optimize federated simulations. By accounting for client dataset heterogeneity, hardware diversity, and communication efficiency
% through its model-based client placement, \emph{optimal concurrency} estimator, and partial intra-node aggregation
, 
\pollen effectively distributes workload across workers and halves the execution time compared to \pfl.
\section{Conclusion}\label{sec:conclusion}
In this work, we propose \pollen, a general-purpose system for federated learning simulation capable of executing large-scale experiments in an unprecedently short time compared to other simulation frameworks. 
We show in our evaluation that \pollen can execute FL experiments with an improvement of days in \emph{medium-scale} settings, weeks in \emph{large-scale} settings, and months in \emph{very large-scale} settings by efficiently utilizing computational resources.
\pollen will enable researchers to experiment in challenging large-scale settings with millions of clients because it benefits grow proportionally to the population size and the hardware available.
Thus, we recommend that \pollen design choices become the de facto standard in federated learning research.

\begin{table}[ht]
    \centering
    \caption{Multi-node comparison between \pollen and \pfl on the two benchmark datasets used by \citet{pfl_paper,pfl_wandb_report}. \pollen outperforms \pfl in all but the smallest CIFAR-10 setting. Even though CIFAR-10 has balanced clients and thus does not suffer from the dataset heterogeneity concerns listed above, the inability of \pfl to account for hardware heterogeneity and the different communication bandwidth between intra-machine and inter-machine scenarios makes it ineffective when scaling to large cohorts.
    %These effects are even more pronounced for the naturally partitioned FLAIR dataset where \pillen executes the \num{2000} client simulation \emph{twice} as fast as \pfl.
    }
    \begin{tabular}{rcc}
        \toprule
         \textbf{Task} & \textbf{\pollen} & \textbf{PFL} \\
         (cohort size, rounds) & [min] & [min] \\
        \hline
        \textbf{CIFAR10} ($50$, $1500$) & 10.7 $\pm$ 0.1 & \textbf{6.40 $\pm$ 0.18} \\
        \textbf{CIFAR10} ($500$, $500$) & \textbf{8.75 $\pm$ 0.04} & 10.08 $\pm$ 0.24 \\
        \textbf{CIFAR10} ($5000$, $250$) & \textbf{29.09 $\pm$ 0.05} & 44.19 $\pm$ 1.17 \\
        \textbf{CIFAR10} ($50000$, $100$) & \textbf{108.3 $\pm$ 0.4} & 172.52 $\pm$ 1.16 \\
        \hline
        \textbf{FLAIR} ($200$, $500$) & \textbf{45.5 $\pm$ 0.1} & 	59.82 $\pm$ 0.90 \\
        \textbf{FLAIR} ($2000$, $50$) & \textbf{25.5 $\pm$ 0.2} & 54.44 $\pm$ 0.73 \\
        % \textbf{FLAIR} ($20000$, $25$) & $\pm$ & $\ast$ \\
        % \textbf{FLAIR} ($200000$, $10$) & $\pm$ & $\ast$ \\
        \bottomrule
    \end{tabular}
    % \vskip +5pt
    \label{tab:pfl_comparison}
% \vspace{-0.4cm}
\end{table}

\clearpage

\bibliographystyle{ACM-Reference-Format}
\bibliography{example_paper}

\clearpage
\appendix
\section{Supplementary Materials}
We provide a series of supplementary material which support the narrative of our text~(i.e., \cref{app:importance_large_scale,app:client_execution,app:gpu_schedulers}), detail additional system design choices in \cref{app:system_details}, and provide greater detail for experimental settings in \cref{app:experimental_settings} and \cref{tab:worker_counts}. Most importantly, in \cref{app:ultra_large_scale_experiments}, we show new results for \emph{ultra large scale} experiments for \emph{Speech Recognition} and \emph{Masked Language Modelling}, which we had to remove from the main text as other frameworks would have taken too long to complete training.

\begin{figure*}[t]
    \centering
    \includegraphics[height=1.0\columnwidth]{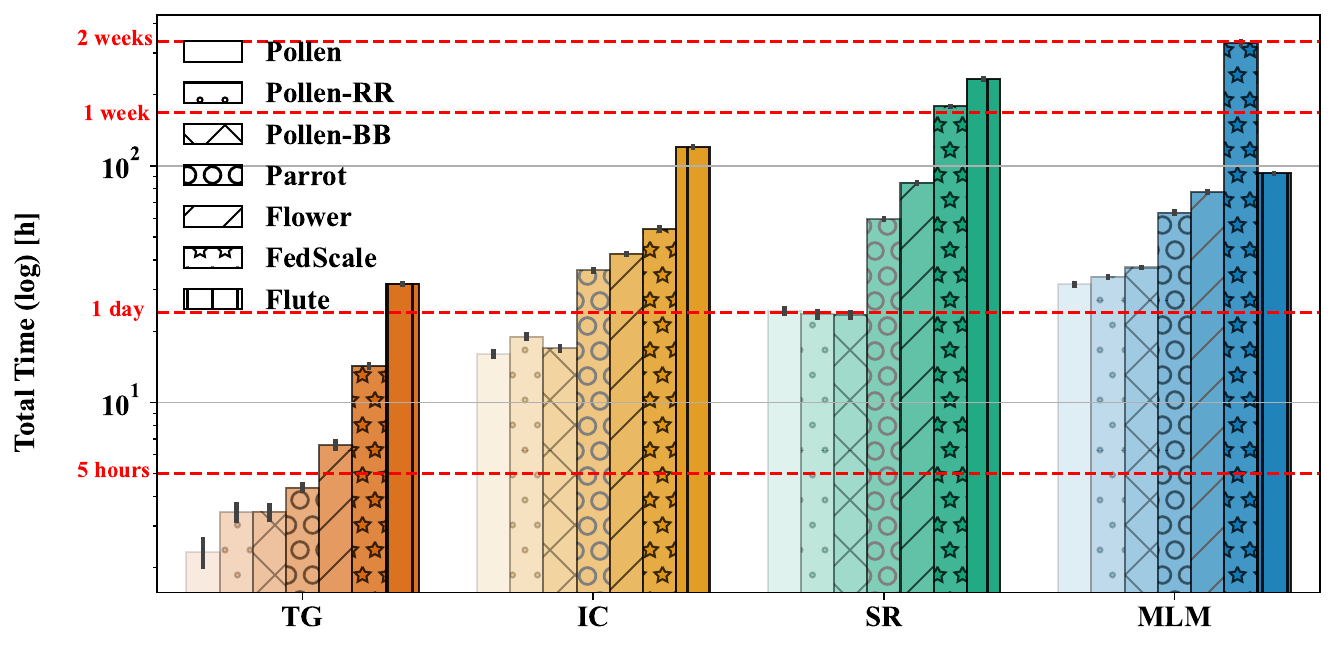}
    \captionsetup{font=small,labelfont=bf}
    \caption{Multi-node comparisons of \pollen against alternative placement methods and frameworks for medium-scale settings. The learning-based placement of \pollen allows it to outperform alternatives in all cases by balancing workload across GPUs.}
    \label{app:fig:multinode_complete}
    \Description[Rebuttal Figure 1]{Multi-node comparisons of \pollen against alternative placement methods and frameworks for medium-scale settings. The learning-based placement of \pollen allows it to outperform alternatives in all cases by balancing workload across GPUs. The different tasks are represented on the x-axis in different colors, while the frameworks are shown in different textures.}
\end{figure*}

\begin{figure*}[t]
    \centering
    \includegraphics[height=1.0\columnwidth]{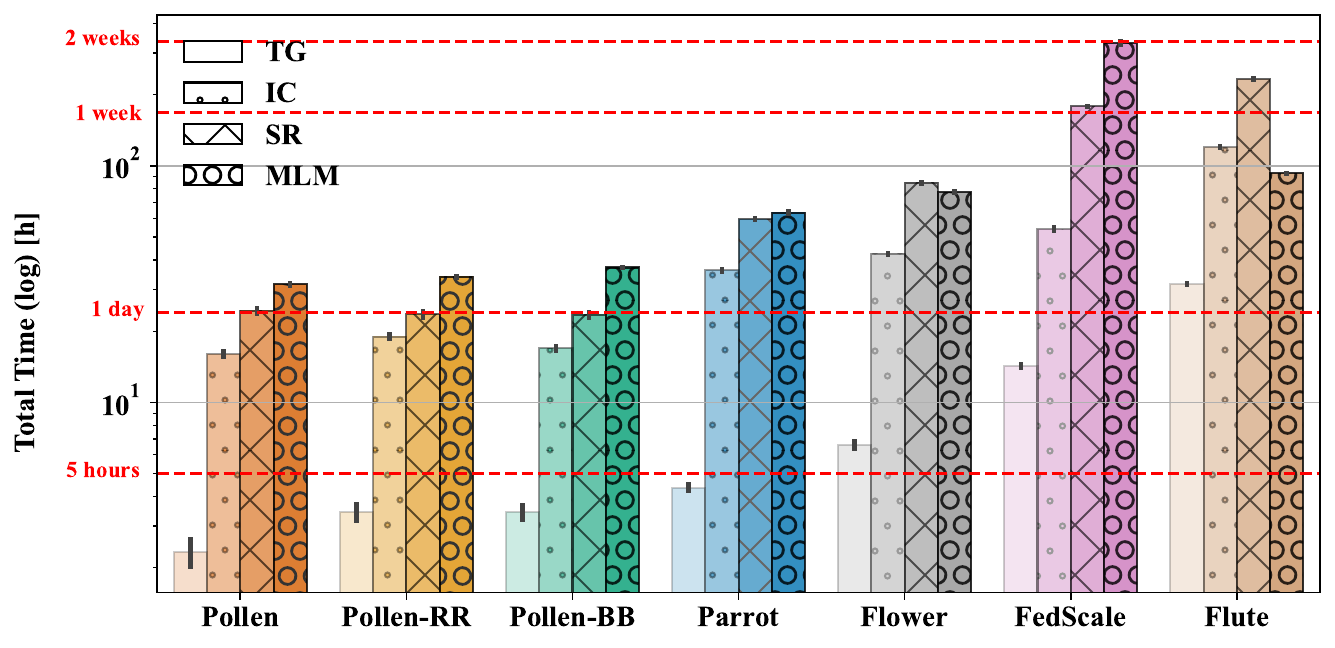}
    \captionsetup{font=small,labelfont=bf}
    \caption{Multi-node comparisons of \pollen against alternative placement methods and frameworks for medium-scale settings. The learning-based placement of \pollen allows it to outperform alternatives in all cases by balancing workload across GPUs.}
    \label{app:fig:multinode_complete_non_inv}
    \Description[Rebuttal Figure 3]{Multi-node comparisons of \pollen against alternative placement methods and frameworks for medium-scale settings. The learning-based placement of \pollen allows it to outperform alternatives in all cases by balancing workload across GPUs. The different frameworks are represented on the x-axis in different colors, while the tasks are shown in different textures.}
\end{figure*}

\begin{figure*}[t]
    \centering
    \includegraphics[height=1.0\columnwidth]{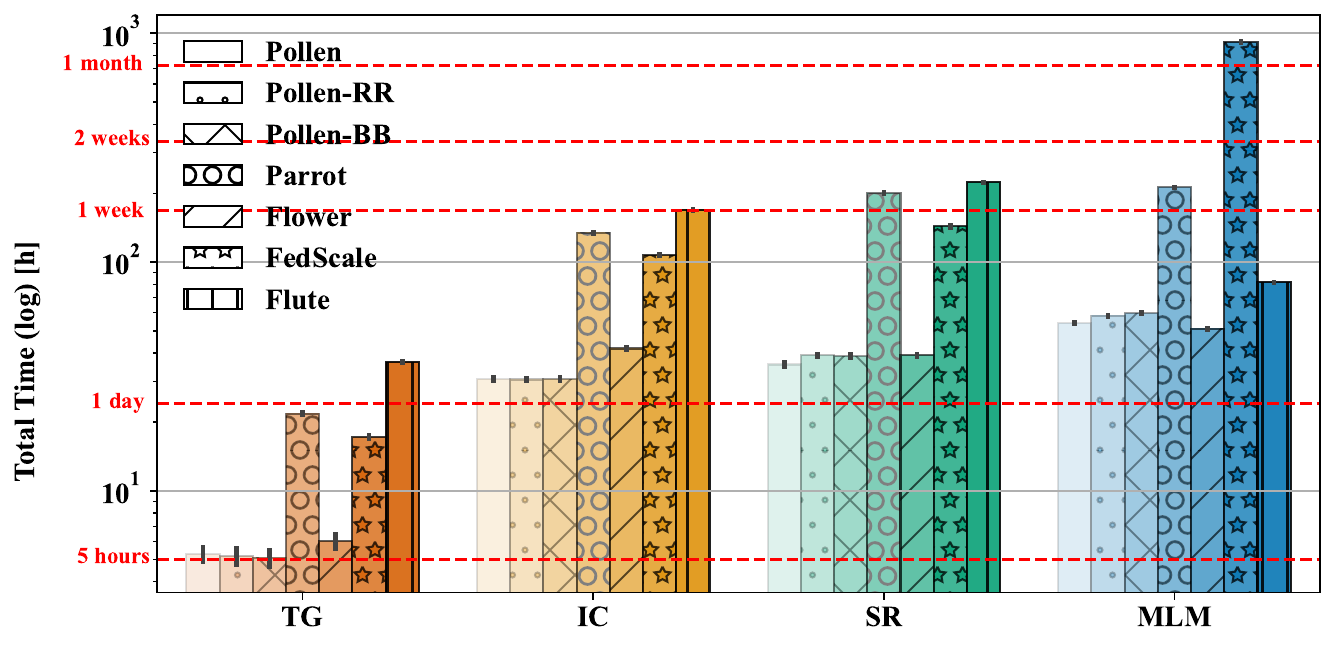}
    \captionsetup{font=small,labelfont=bf}
    \caption{Single-node comparisons of \pollen against alternative placement methods and frameworks for medium-scale settings. The efficient implementation of learning-based placement allows it to match alternatives despite no heterogeneity being present in the system.}
    \label{app:fig:singlenode_complete}
    \Description[Rebuttal Figure 2]{Single-node comparisons of \pollen against alternative placement methods and frameworks for medium-scale settings. The efficient implementation of learning-based placement allows it to match alternatives despite no heterogeneity being present in the system. The different tasks are represented on the x-axis in different colors, while the frameworks are shown in different textures.}
\end{figure*}

\begin{figure*}[t]
    \centering
    \includegraphics[height=1.0\columnwidth]{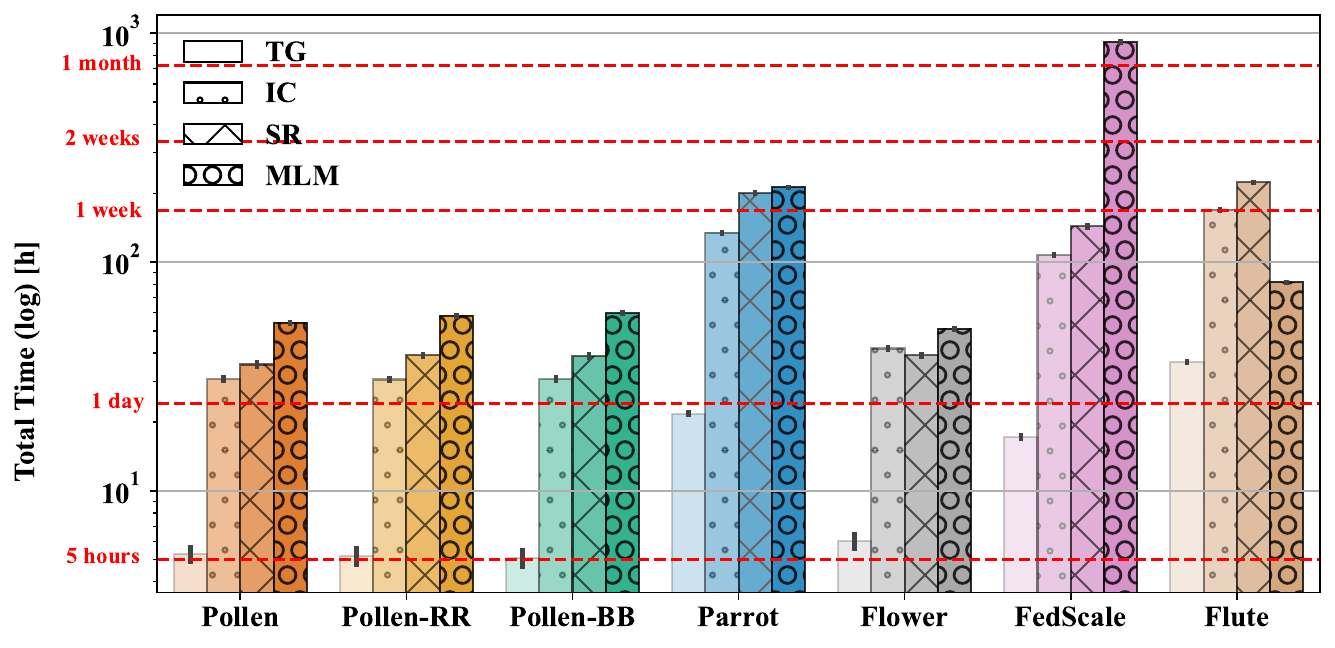}
    \captionsetup{font=small,labelfont=bf}
    \caption{Single-node comparisons of \pollen against alternative placement methods and frameworks for medium-scale settings. The efficient implementation of learning-based placement allows it to match alternatives despite no heterogeneity being present in the system.}
    \label{app:fig:singlenode_complete_non_inv}
    \Description[Rebuttal Figure 4]{Single-node comparisons of \pollen against alternative placement methods and frameworks for medium-scale settings. The efficient implementation of learning-based placement allows it to match alternatives despite no heterogeneity being present in the system. The different frameworks are represented on the x-axis in different colors, while the tasks are shown in different textures.}
\end{figure*}

\begin{figure*}[t]
    \centering
    \includegraphics[height=1.0\columnwidth]{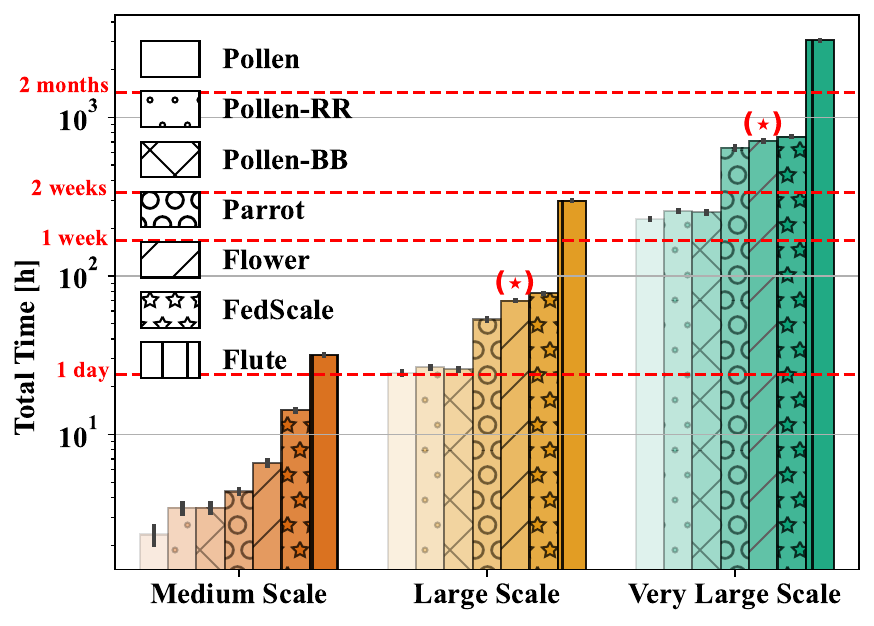}
    \captionsetup{font=small,labelfont=bf}
    \caption{Scalability comparison of \pollen against alternative placement methods and frameworks for Text Generation. The learning-based placement of \pollen maintains an advantage across scales. Asterisks indicate training failures.}
    \label{app:fig:tg_scalability_complete}
    \Description[Rebuttal Figure 5]{Scalability comparison of \pollen against alternative placement methods and frameworks for Text Generation. The learning-based placement of \pollen maintains an advantage across scales. Asterisks indicate training failures.}
\end{figure*}

\begin{figure*}[t]
    \centering
    \includegraphics[height=1.0\columnwidth]{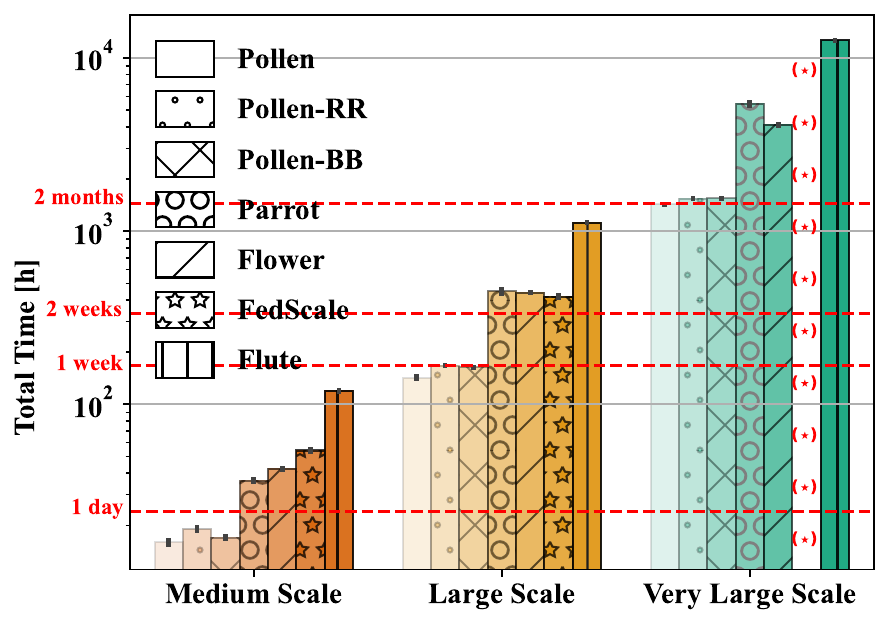}
    \captionsetup{font=small,labelfont=bf}
    \caption{Scalability comparison of \pollen against alternative placement methods and frameworks for Image Classification. The learning-based placement of \pollen maintains an advantage across scales. Asterisks indicate training failures.}
    \label{app:fig:ic_scalability_complete}
    \Description[Rebuttal Figure 6]{Scalability comparison of \pollen against alternative placement methods and frameworks for Image Classification. The learning-based placement of \pollen maintains an advantage across scales. Asterisks indicate training failures.}
\end{figure*}

\begin{figure*}[t]
    \centering
    \includegraphics[height=1.0\columnwidth]{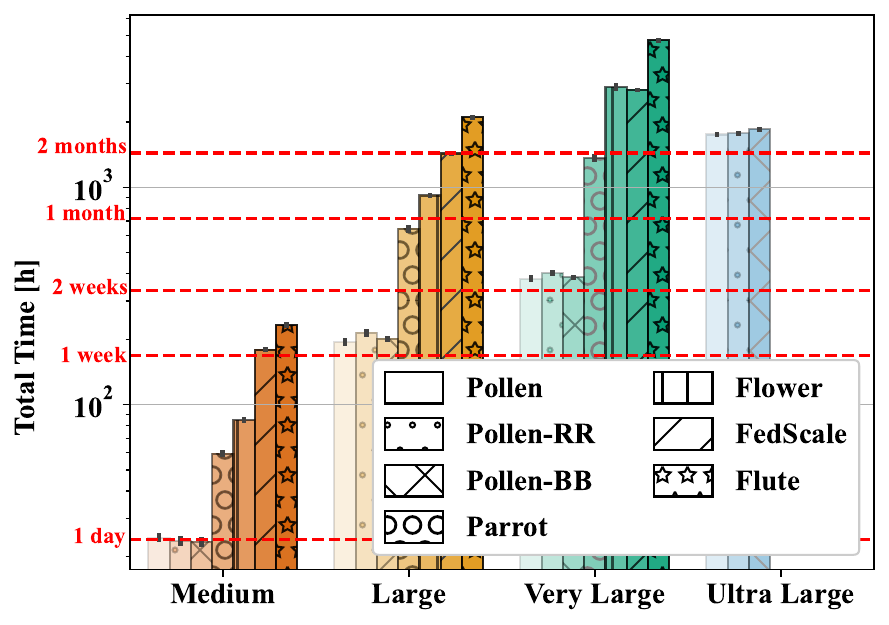}
    \captionsetup{font=small,labelfont=bf}
    \caption{Scalability comparison of \pollen against alternative placement methods and frameworks for Speech Recognition. The learning-based placement of \pollen maintains an advantage across scales, with only the \pollen versions able to complete the largest scale.}
    \label{app:fig:sr_scalability_complete}
    \Description[Rebuttal Figure 7]{Scalability comparison of \pollen against alternative placement methods and frameworks for Speech Recognition. The learning-based placement of \pollen maintains an advantage across scales, with only the \pollen versions able to complete the largest scale.}
\end{figure*}

\begin{figure*}[t]
    \centering
    \includegraphics[height=1.0\columnwidth]{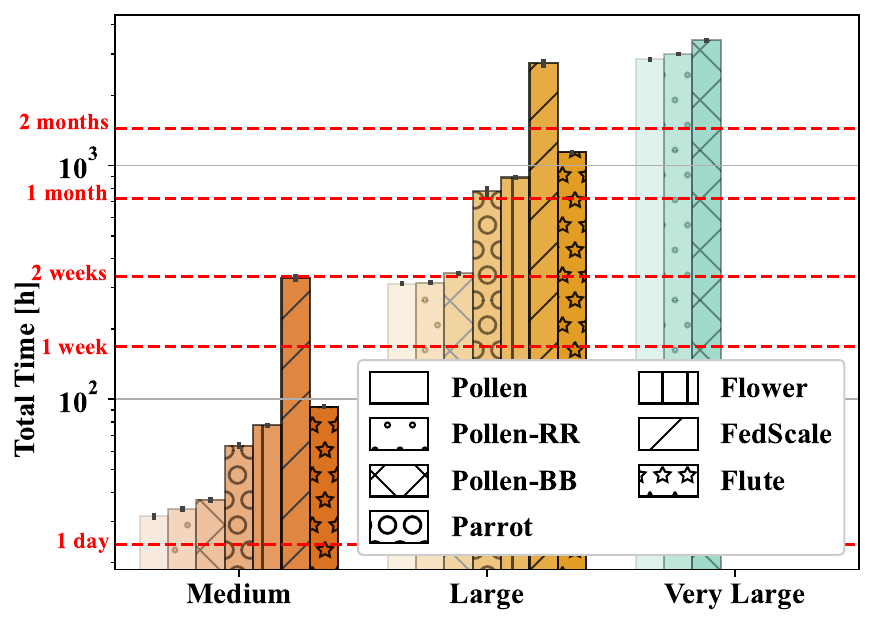}
    \captionsetup{font=small,labelfont=bf}
    \caption{Scalability comparison of \pollen against alternative placement methods and frameworks for Masked Language Modelling. The learning-based placement of \pollen maintains an advantage across scales, with only the \pollen versions able to complete the largest scale.}
    \label{app:fig:mlm_scalability_complete}
    \Description[Rebuttal Figure 8]{Scalability comparison of \pollen against alternative placement methods and frameworks for Masked Language Modelling. The learning-based placement of \pollen maintains an advantage across scales, with only the \pollen versions able to complete the largest scale.}
\end{figure*}

In addition to the latter, we also show more extensive experimental results showcasing \pollen's performance improvement compared to other frameworks and different placement strategies for the \emph{single-node} (\cref{app:fig:singlenode_complete,app:fig:singlenode_complete_non_inv}), \emph{multi-node} (\cref{app:fig:multinode_complete,app:fig:multinode_complete_non_inv}) and scalability (\cref{app:fig:ic_scalability_complete,app:fig:sr_scalability_complete,app:fig:tg_scalability_complete,app:fig:mlm_scalability_complete}).

More benchmarks regarding resource utilization show, in \cref{app:tab:gpu_util,app:tab:gpu_mem}, that even in the setting with limited optimization gaps, i.e.~\emph{single-node}, \pollen produces high GPU utilization and memory allocation.

At last, we show numerical results (\cref{app:tab:fedmedian_agg_time}) that can help the reader set a reference for the impact that full aggregation may have on the training time.
In \cref{app:tab:idle_time}, we report the estimated idle time for experiments in the \emph{multi-node} setting, comparing \pollen against other placement strategies.

\subsection{Further Experimental Details}\label{app:experimental_settings}

This subsection provides additional hyperparameters excluded from the main text due to space constraints. The cohorts sampled for each round of our experimentation were generated using a pseudo-random number generator (PRNG) from the default Python library seeded with $1337$. The same seed was used for all random behaviors. Our Python version was $3.9.17$, with our environment being constructed via poetry and thus reproducible. The PyTorch version we used was $1.12.1$, running on CUDA version $11.3$. For the transformers library, we used version $4.34.0$.

\noindent\textbf{Image Classification (IC)} We use a batch size of \num{20} samples.
The local optimizer clients use during local training is SGD with learning rate $\eta=0.05$, momentum $m=0.9$, and weight decay $\tau=5\times10^{-4}$. 

\noindent\textbf{Speech Recognition (SR)}  We use a batch size of \num{20} samples.
The clients use the same local optimizer as the \emph{image classification} task.

\noindent\textbf{Text Generation (TG)} 
We follow the LEAF experimental configuration and use a batch size of \num{4} samples.
The clients use the same local optimizer as the \emph{image classification} task, except for the learning rate being $\eta=0.8$.

\noindent\textbf{Masked Language Modelling (MLM)} We use a batch size of \num{20} samples. The local optimizer clients use is Adam with learning rate $\eta=4\times10^{-5}$, $(\beta_1,\beta_2)=(0.99, 0.999)$ and weight decay $\tau=5\times10^{-2}$.

\noindent\textbf{Statistics Estimation} To facilitate the speed of framework comparisons, we gathered statistics throughout \num{100} rounds of execution for all frameworks and extrapolated to \num{5000} rounds as a commonly used experimental setting in FL research.

\noindent\textbf{Placement Policy Comparisons} We used training times taken from \emph{Round-Robin} experiments as our baseline data for the comparison since they were unbiased by an intelligent placement procedure. To simulate the real idle times of the workers, we used the statistics gathered from these experiments to estimate the real load following the decision made by our \emph{Learning-Based} placement procedure.

\begin{table}
    \centering
    \caption{The \emph{number of concurrent processes} executed for the tasks on each GPU type we used. The header codes identify the tasks.}
    \begin{tabular}{rcccc}
        \toprule
        \textbf{GPU type} & \textbf{TG} & \textbf{IC} & \textbf{SR} & \textbf{MLM} \\
        \hline
        \emph{Nvidia A40} & 33 & 14 & 21 & 14 \\
        \emph{Nvidia 2080 Ti} & 10 & 4 & 7 & 3 \\
        \bottomrule
    \end{tabular}
    \label{tab:worker_counts}
\end{table}

\subsection{Very and Ultra Large Scale Experiments}\label{app:ultra_large_scale_experiments}
\begin{figure}[ht]
    \centering
    \noindent\includegraphics[height=0.7\columnwidth]{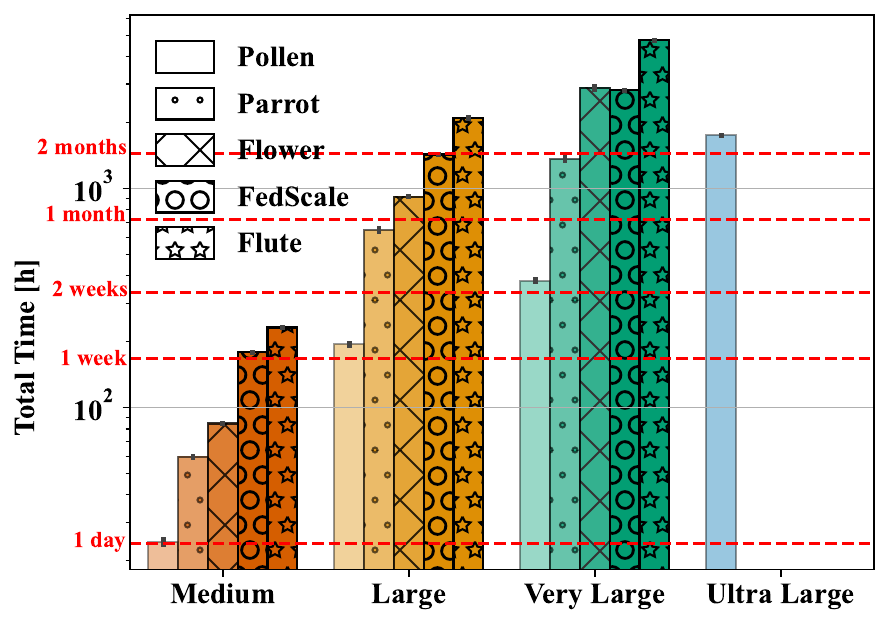}
    \captionsetup{font=small,labelfont=bf}
    \caption{\emph{Ultra-large} scalability comparisons for \emph{Speech Recognition}. \pollen is the only framework capable of simulating \num{10000} clients per round for \num{5000} rounds within a reasonable time for this task. Remarkably, it takes as long, or less, to complete \emph{ultra-large} scale experiments with \pollen as it takes to complete merely \emph{very-large} ones for the other frameworks.}
    \label{app:fig:Scale_SR}
    \Description[Appendix Figure 1]{\emph{Ultra-large} scalability comparisons for \emph{Speech Recognition}. \pollen is the only framework capable of simulating \num{10000} clients per round for \num{5000} rounds within a reasonable time for this task. Remarkably, it takes as long, or less, to complete \emph{ultra-large} scale experiments with \pollen as it takes to complete merely \emph{very-large} ones for the other frameworks.}
% \vspace{-0.4cm}
\end{figure}

\begin{figure}[ht]
    \centering
    \noindent\includegraphics[height=0.7\columnwidth]{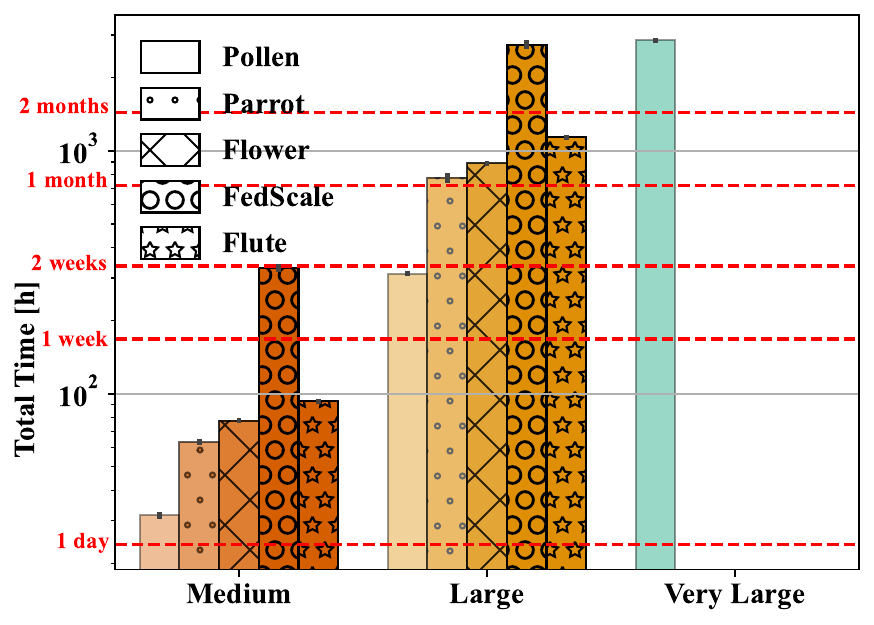}
    \captionsetup{font=small,labelfont=bf}
    \caption{\emph{Very-large} scalability comparisons for \emph{Masked Language Modelling}. \pollen is the only framework capable of simulating \num{10000} clients per round for \num{5000} rounds within a reasonable time for this task. Similarly to the \emph{Speech Recognition} task, its execution time for \emph{very-large} scale experiments is on the same order as that of merely \emph{large} ones for the other frameworks.}
    \label{app:fig:Scale_MLM}
    \Description[Appendix Figure 2]{\emph{Very-large} scalability comparisons for \emph{Masked Language Modelling}. \pollen is the only framework capable of simulating \num{10000} clients per round for \num{5000} rounds within a reasonable time for this task. Similarly to the \emph{Speech Recognition} task, its execution time for \emph{very-large} scale experiments is on the same order as that of merely \emph{large} ones for the other frameworks.}
% \vspace{-0.4cm}
\end{figure}

\Cref{app:fig:Scale_SR} and \cref{app:fig:Scale_MLM} showcase experiments using \num{10000} clients per round for \num{5000} rounds for the \emph{Speech Recognition} and \emph{Masked Language Modelling Tasks}. These were excluded from the main text due to time constraints caused by the other frameworks. They showcase the benefits of \pollen on two very challenging tasks beyond the bounds of possibility for \flower, \fedscale, \flute, and \parrot. Thus, they provide strong evidence for the future-proof nature of \pollen's design and further demonstrate its status as the only reasonable choice for genuine \emph{large-scale} research.

\subsection{System Design Details}\label{app:system_details}

In this section, we want to emphasize two elements of our \pollen design. The first is the extent to which Partial Aggregation can reduce communication costs in large-scale settings. Specifically, we argue that in the absence of degenerate cases, such as scenarios in which the number of clients is so great that aggregation never completes, partial aggregation allows nodes to perform constant-size communication to the server. By constant size, we mean the number of rounds and the number of clients per round, as the model is assumed to be fixed. Of course, this only holds for associative strategies.

The first is the difficulties posed by systems that do not allow automatic concurrency estimation. Indeed, if manual configurations are required, then an entire separate experiment to evaluate the resource consumption of the simulation is necessary for every attempted configuration, thus increasing the overall time of the research. Furthermore, this process must be repeated for every task and re-started whenever the hardware configuration changes.

\subsection{The importance of Large Scale FL}\label{app:importance_large_scale}

As edge hardware's ubiquity and compute power increase and privacy regulations and concerns spread worldwide~\citep{united-nations}, data-hungry models require FL to reach global scales. Despite its potential to harvest data and computing power from millions of devices, we know surprisingly little about how well FL solutions may scale to such numbers due to the lack of experiments, and although some characteristics are common to all FL systems, some are only noticeable at large scales. 
At the same time, production systems at Google~\citep{ScaleAndSystemDesign}, Meta~\citep{PAPAYA}, and Apple~\citep{AppleFL} orchestrate hundreds of millions of devices, with thousands available per round, casting doubt on the validity of small-scale research. Thus, unaffiliated researchers need a far more efficient simulator to maintain the relevance of their work to production systems. Due to limited FL framework capability 

\subsection{Additional Challenges of Simulating FL}\label{app:client_execution}

Several factors that affect the actual training time of a client discussed in \cref{subsec:hardware_heterogeneity} are worth expanding upon.
First, different datasets have extremely divergent pre-processing pipelines. For example, image datasets may require samples to be loaded from disk and transformed, while language datasets may store features in RAM. Second, the CPU has a significant impact on the dataloading efficiency of client execution and interacts with the concurrency supported by the GPU. For example, as the number of concurrent CPU jobs grows beyond the number of CPU cores, the duration of data loading time grows proportionally and is strongly affected by the scheduling of the Operating System.
This can act as a bottleneck to increasing the number of workers available.
Third, since clients in cross-device FL are relatively small jobs, they allow significantly greater flexibility than large jobs in traditional batch-processing systems.

\subsection{Limitations of General GPU Schedulers for FL}\label{app:gpu_schedulers}
General limitations of GPU scheduling for ML tasks are also worth mentioning.
Job scheduling on a GPU cluster for traditional ML tasks is a well-studied problem~\citep{GpuSchedulingSurvey, Gandiva, CODA}.
For example, Gandiva~\citep{Gandiva} schedules large jobs on such a cluster by profiling the rate at which they process mini-batches.
At the same time, CODA~\citep{CODA} attempts to balance the CPU assignment of GPU jobs to optimize data loading.
However, both rely on the assumption that ML tasks are highly repetitive and run long enough to be effectively optimized.
As we will show in \cref{sec:sim_fl}, client dataset sizes in FL are much smaller than a typical ML job and more difficult to profile.
Furthermore, client dataset sizes are highly skewed, making round durations unpredictable.
Together, these two considerations make the heuristics of existing GPU scheduling systems insufficient for FL\@.
This insufficiency drives us to propose an FL-specific solution rather than plugging in an existing one.

\subsection{Infrastructure for Large-Scale Simulation}\label{app:systems_engineering}
We optimize the intra-node execution of our system to minimize memory allocations and data movement on a given machine by preferring in-place operations.
For example, each worker process maintains the model parameters used for training in shared memory, to which we write the initial model before training each client.
Partially aggregated models on the workers and nodes use the same technique.

% \clearpage

\begin{table*}[t]
    \centering
    \caption{Framework comparison in terms of \textit{percentual GPU utilization} during a single-round in the \textit{single-node} setting. The metric is obtained using \texttt{nvidia-smi} and averaged over the time elapsed to complete the round. The second round has been chosen to exclude initialization effects. \pollen leverages efficient multiprocessing on a single GPU, finishing with the highest or second-highest utilization in all tasks. While \flower does well in this setting, it cannot be generalized to multiple GPUs since it cannot account for hardware diversity.}
    \begin{tabular}{rccccc}
    \toprule
     \textbf{Task} & \textbf{FedScale} & \textbf{Flower} & \textbf{Flute} & \textbf{Parrot} & \textbf{\pollen} \\
    \hline
    \textbf{IC} & 19.023 $\pm$ 0.002 & 66.987 $\pm$ 0.005 & 13.7509 $\pm$ 0.0003 & 16.7259 $\pm$ 0.0003 & \textbf{92.415 $\pm$ 0.003} \\
    \textbf{MLM} & 4.54930 $\pm$ 6e-05 & \textbf{83.314 $\pm$ 0.004} & 22.2825 $\pm$ 0.0006 & 18.6346 $\pm$ 0.0003 & 80.563 $\pm$ 0.003 \\
    \textbf{SR} & 4.3110 $\pm$ 0.0002 & 19.666 $\pm$ 0.001 & 4.84215 $\pm$ 9e-05 & 3.15926 $\pm$ 6e-05 & \textbf{21.344 $\pm$ 0.002} \\
    \textbf{TG} & 35.28 $\pm$ 0.02 & \textbf{86.40 $\pm$ 0.02} & 21.987 $\pm$ 0.001 & 29.8795 $\pm$ 0.0007 & 85.81 $\pm$ 0.03 \\
    \bottomrule
    \end{tabular}
    \captionsetup{font=small,labelfont=bf}
    \label{app:tab:gpu_util}
\end{table*}

% \FloatBarrier

\begin{table*}[t]
    \centering
    \caption{Framework comparison in terms of \textit{percentual VRAM allocation} during a single-round in the \textit{single-node} setting. The metric is obtained using \texttt{nvidia-smi} and averaged over the time elapsed to complete the round. The second round has been chosen to exclude initialization effects. \pollen achieves the highest VRAM allocation in three out of four, while the single-worker frameworks \flute and \parrot fail to use the GPU to its maximum potential.}
    \begin{tabular}{rccccc}
    \toprule
     \textbf{Task} & \textbf{FedScale} & \textbf{Flower} & \textbf{Flute} & \textbf{Parrot} & \textbf{\pollen} \\
    \hline
    \textbf{IC} & 89.9367 $\pm$ 0.0003 & 89.658 $\pm$ 0.002 & 5.57037 $\pm$ 3e-05 & 6.70$\dotsc$ $\pm$ $\sim0$ & \textbf{93.26503 $\pm$ 2e-05} \\
    \textbf{MLM} & 7.57$\dotsc$ $\pm$ $\sim0$ & \textbf{96.559161 $\pm$ 2e-06} & 5.04664 $\pm$ 6e-05 & 7.45$\dotsc$ $\pm$ $\sim0$ & 67.602 $\pm$ 0.001 \\
    \textbf{SR} & 48.3002 $\pm$ 0.0001 & 91.100991 $\pm$ 5e-06 & 4.490179 $\pm$ 1e-06 & 4.57$\dotsc$ $\pm$ $\sim0$ & \textbf{95.653381 $\pm$ 3e-06} \\
    \textbf{TG} & 75.975 $\pm$ 0.002 & 92.64$\dotsc$ $\pm$ $\sim0$ & 2.887706 $\pm$ 1e-06 & 2.91$\dotsc$ $\pm$ $\sim0$ & \textbf{98.42697 $\pm$ 1e-05} \\
    \bottomrule
    \end{tabular}
    \captionsetup{font=small,labelfont=bf}
    \label{app:tab:gpu_mem}
\end{table*}

\begin{table*}[t]
    \centering
    \caption{The duration of the aggregation at the server using \textit{FedMedian} for all the tasks. Three different values for the number of clients/models have been used, i.e., 10, 100, 1000, to show how such a duration scales. Note that the different tasks are representative of the models' scales: TG weighs 3.28MB, IC weighs 26.45MB, MLM weighs 60.37MB, and SR weighs 85.14MB.}
    \begin{tabular}{rcccccc}
    \toprule
    & \textbf{\#Clients per round} & \textbf{TG} [s] & \textbf{IC} [s] & \textbf{MLM} [s] & \textbf{SR} [s] \\ 
    \hline
    \textit{Full} &\textbf{10} & 0.17 & 1.44 & 1.47 & 4.78 \\
    \textit{Aggregation} &\textbf{100} & 1.58 & 12.76 & 18.58 & 46.41 \\
    \textit{Time} &\textbf{1000} & 30.47 & 167.66 & 352.45 & 812.01 \\
    \bottomrule
    \end{tabular}
    \captionsetup{font=small,labelfont=bf}
    \label{app:tab:fedmedian_agg_time}
\end{table*}

\begin{table*}[ht]
    \centering
    \caption{Placement efficiency for \pollen, \emph{Na\"{\i}ve Round-Robin} and \emph{Batches-Based}. \pollen minimizes GPU idle time across all tasks.}
    \begin{tabular}{rccc}
    \toprule
        \textbf{Idle time [s]} & \textbf{\pollen} & \textbf{RR} & \textbf{BB} \\
        \hline
        \textbf{SR} & \textbf{1082 $\pm$ 9} & 1469 $\pm$ 12 & 1456 $\pm$ 10 \\
        \textbf{TG} & \textbf{7386 $\pm$ 8} & 14259 $\pm$ 8 & 13144 $\pm$ 4 \\
        \textbf{IC} & \textbf{34590 $\pm$ 26} & 57681 $\pm$ 25 & 58376 $\pm$ 9 \\
        \textbf{MLM} & \textbf{71193 $\pm$ 54} & 121620 $\pm$ 32 & 118962 $\pm$ 12 \\
        \bottomrule
    \end{tabular}
    \label{app:tab:idle_time}
    \label{tab:idle_time_comparison}
\end{table*}

\end{document}